\pdfoutput=1

		\documentclass{isprs}
		\usepackage{hyperref}
		\usepackage{subfigure} 

		\usepackage[utf8]{inputenc}
	
		\usepackage{varioref} 
	 
		\usepackage{natbib} 
	\usepackage{enumitem}
		\usepackage{graphicx}
		\DeclareGraphicsExtensions{.pdf,.jpg,.png}

		\usepackage[usenames,dvipsnames]{color}
	\usepackage{amsmath}
		\usepackage{algorithm2e}
		\usepackage{microtype}  
		\usepackage{SIunits} 
		\usepackage{listings}
		\usepackage{tabularx} 
		\usepackage{setspace}

		\newcommand{\myimage}[3]
					{
					\begin{figure} [h!]
						\begin{center}
							\includegraphics[width=\linewidth,keepaspectratio]{#1}
							\caption{#2}  
							\label{#3}
							\end{center}
					\end{figure} 
					}

		\newcommand{\myimageHL}[4]
		{
			\begin{figure} [ht!]
				\begin{center}
					\includegraphics[width= #4 \linewidth ,keepaspectratio]{#1}
					\caption{#2}  
					\label{#3}
				\end{center}
			\end{figure} 
		}

		\newcommand{\myimageFullPageWidth}[3]
								{
								\begin{figure*}[ht]
									\begin{center}
										\includegraphics[width=\textwidth,keepaspectratio ]{#1}
										\caption{#2}  
										\label{#3} 
										\end{center}
								\end{figure*} 
								}

		\newcommand{\myimageInline}[2]
					{
					\myimage{#1}{#2}
					}	 
		
		\newcommand{\myemph}[1]{\emph{#1}}

	\usepackage{rotating}

\title{A state of the art of urban reconstruction: street, street network, vegetation, urban feature}
\author{Rémi Cura  $^{A}$, Julien Perret $^A$, Nicolas Paparoditis  $^A$}

\address{ $^A$  Université Paris-Est, IGN, SRIG, COGIT \& MATIS, 73 avenue de Paris, 94160 Saint Mandé, France\\
	first\_name.last\_name@ign.fr
	}

\begin{document}
 



\abstract{
World population is raising, especially the part of people living in cities.
With increased population and complex roles regarding their inhabitants and their surroundings, cities concentrate difficulties for design, planning and analysis.
These tasks require a way to reconstruct/model a city.

Traditionally, much attention has been given to buildings reconstruction, yet an essential part of city were neglected: streets.
Streets reconstruction has been seldom researched.
Streets are also complex compositions of urban features, and have a unique role for transportation (as they comprise roads).
We aim at completing the recent state of the art for building reconstruction~\citep{Musialski2012a} by considering all other aspect of urban reconstruction. 
We introduce the need for city models (Sec. \vref{sota.intro}).
Because reconstruction always necessitates data, we first analyse which data are available (Sec. \vref{sota.input_data}).
We then expose a state of the art of street reconstruction (Sec. \vref{sota.street}), street network reconstruction (Sec. \vref{sota.street_network}), urban features reconstruction/modelling (vegetation (Sec. \vref{sota.vegetation}), and urban objects reconstruction/modelling (Sec. \vref{sota.urban_feature}).
 
Although reconstruction strategies vary widely, we can order them by the role the model plays, from data driven approach, to model-based approach, to inverse procedural modelling and model catalogue matching.
The main challenges seems to come from the complex nature of urban environnement and from the limitations of the available data.

Urban features have strong relationships, between them, and to their surrounding, as well as in hierarchical relations.
Procedural modelling has the power to express these relations, and could be applied to the reconstruction of urban features via the \emph{Inverse Procedural Modelling paradigm}. 
}

\maketitle 


\myimageHL{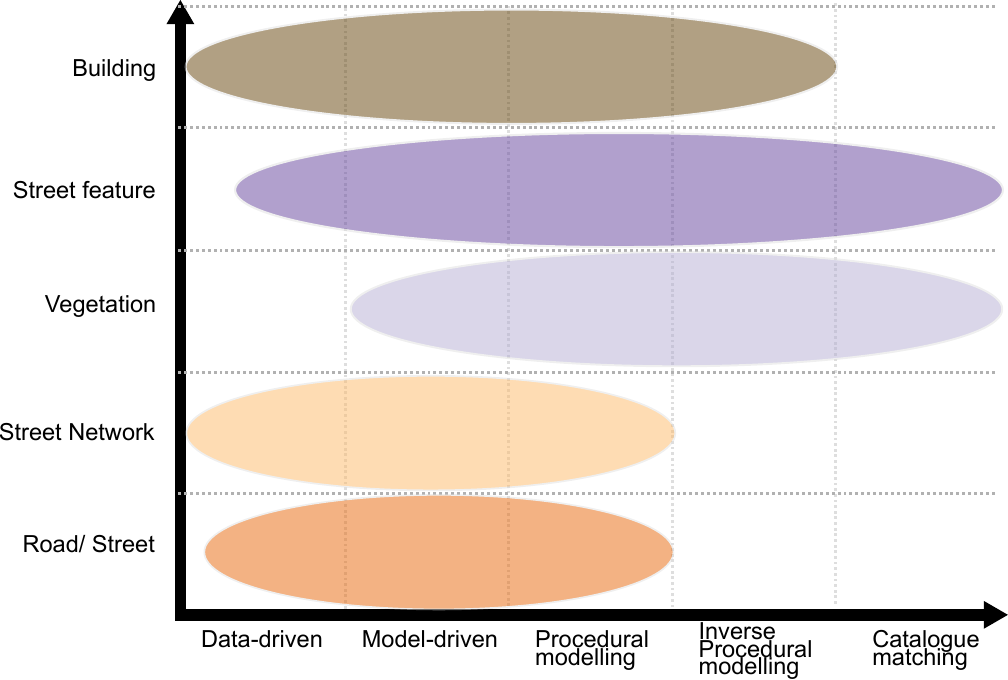}{A subjective transverse classification of reconstruction methods by role of model.}{sota.fig:transverse_reconstruction}{1}
 

\section{Introduction}
	\label{sota.intro}
 
	\subsection{Context}
			
		\subsubsection{Total population living in cities is growing}
			
			World population is increasing fast. 
			A recent survey \citep{UnitedNations2012} shows that 52\% of Mankind already lives in urban area .
			 
			These urban areas are expected to absorb more than the demographic augmentation, with new cities reaching the million of inhabitants every year in Africa and Asia.
			
			The cities not only grow by number of inhabitants but also by the area they occupy. 
			The urban land use is expected to increase in the order of 100 000's \kilo\squaremetre~in the next decade \citep{Seto2011}.	  
			
		\subsubsection{Tensions are building up}
			
			\paragraph{Demographic pressure. : concentration of population  }
				
				While the number of cities is growing, cities are also getting bigger: 40\% of city inhabitants are living in cities over 1 million inhabitants.				
				The growth of cities is partially absorbed by the constitution of megacities.
				 10\% of world population lives in megacities (23 cities that are bigger than 10 millions), and this should increase to 13.5\%  until 2025.

			\paragraph{Social pressure}
				
				Cities also concentrate inequalities, which are rising in the country where the urbanisation is expected to be the most significant (\cite{OECD2010}, p.37). 
			
			\paragraph{Environmental pressure}
				
				High densities in cities imply careful management of environment of a city. The necessary fluxes (in and out) are massive. Although some of these fluxes are natural, they are also heavily impacted by cities (water, air, heat, etc.).
				
			\paragraph{Crisis management}
				Concentration also makes crisis management much more difficult.
				Natural hazard (flood, earthquake, power cut) have more potent effects when hitting a city as they concern more people, and as the very density and complexity of city infrastructures might leave them more vulnerable. 
				Cities growing very fast may outgrow their infrastructures. 
				
				Moreover, cities importance also makes them more susceptible to human-related hazard (epidemic, toxic dispersion).
		 
		\subsubsection{Need for urbanism and city planning}
			For about one century the field of urbanism has been dedicated to tackle those problems. 
			The new challenges and the change of scale of the problem necessitate new tools.

	\subsection{Stakes}
	
		\subsubsection{Need for city modelling}
			Urbanists traditionally use methodologies from social sciences.
			The advance of computer science and engineering has given them simulations tools to model behaviours and even test planning scenarios.
			Planing is spread across several entities, public or private, as well as the inhabitants.
			This makes communication an important aspect of city planing.
			Moreover, the representation of situations and scenario is essential for the decision process as well as for the elaboration of the planning.
		
		\subsubsection{A new tools: the city model}
			2D maps have been the tool of choice, and can now advantageously be completed by structured 3D city models created from various information.
		
		\subsubsection{Answer part of the needs}
			This new model and the 3D nature brings in turn several new applications (see \citep{Niggeler2009} (fr. and ge.). The Figure \ref{sota.fig.3d_model_usage} (inspired from \citep{Niggeler2009}) gives an overview of some applications for a city model.
		
	\subsection{Applications}
		\myimageHL{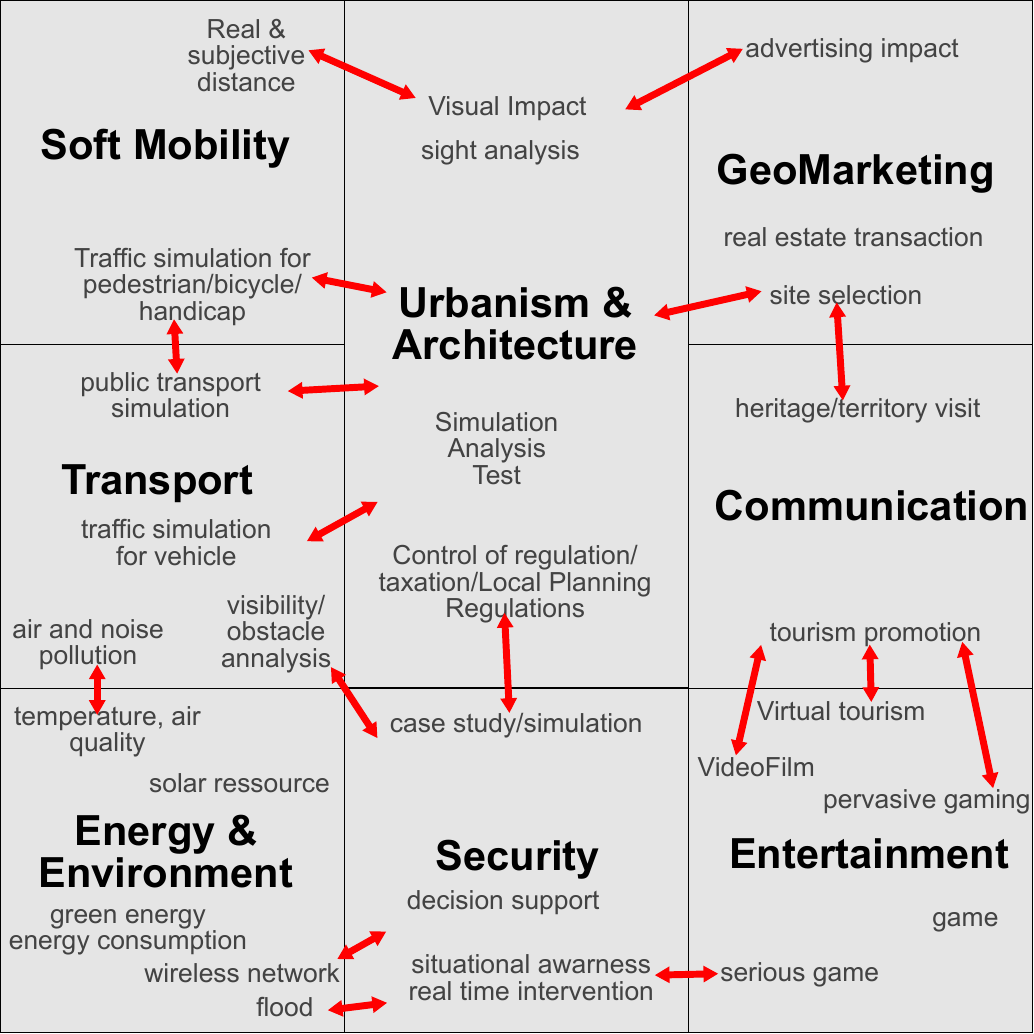}{Example of potential usages of a city model.}{sota.fig.3d_model_usage}{1} 
	
		\subsubsection{Urbanism-related applications} 
			(See Fig. \ref{sota.fig.3d_model_usage})
			City modelling is widely used for urban planning and understanding.
			Having detailed city models is an asset for visualisation and simulation, permitting to test planning scenarios (new and transformation), analysing various impacts and properties (noise, pollution, light propagation, flood, power cut, epidemic, toxic dispersion, water management, temperature), or design transportation system. 
			
			Being a place of spatial and social concentration, a city is very sensible to environment issues.
			Monitoring and simulating air quality~\citep{Moussafir2013}, temperature, wind speed, solar exposition, water cycle and so is important both for social reasons (perceived cleanness, perceived lightness), for energetical reasons (urban heating or cooling), as well as for health (being of high density, cities are more prone to epidemics).
			
				
			Cities models are also used for tourism and communication as a part of the larger Virtual Reality (VR) trend.
			Similarly, digital mapping is used as a simpler VR application, permitting to help in a GPS-based navigation system, or simply browse pictures of the roadside.

		\subsubsection{3D model for entertainment}
			(See Fig. \ref{sota.fig.3d_model_usage})
			The important place the cities have in our lives logically pervades into the collective images used by the entertainment industry. 
			
			Thereby many films pictures real or imagined cities, in particular to support special effects.
			The game industry needs are even bigger, partly because the recent trend toward Massively Multiplayer Online Role-Playing Game (MMORPG).
			Such games often induce massive open cities with believable animated agents.
			Interestingly, the city simulation games are similar in nature to serious training tools used to prepare emergency response, crisis management and police deployment.
	
	\subsection{What is a city}
	Defining properly a city is difficult, as it would involves historical and social criterias.
	In this work, we consider that a city is a densely populated and socially complex place fulfilling certain functions to its surroundings.
	
	\subsection{Challenges}
		Reconstructing 3D city model is vital for many applications that are necessary to manage and plan cities, easing the life of Billions of citizens. 
		
		\subsubsection{Reconstructing a multi-level complex of objects} 
			Some of the reasons that make city modelling particularly challenging can be derived from the definition we used for city:
			A city is a set of connected components interacting with each other. In this way the road network usually influences building placement.
			The social complexity accounts for various uses of space and therefore various types of buildings (constructed for various usage and through the time).
			The dense population use multi modal transportation networks which share part of space (e.g. bicycle and cars).
			
			Such a layered nature incite us to decompose the city reconstruction problem into connected problems:
			 the reconstruction of buildings (which we defer to \citep{Musialski2012a})
			 , the reconstruction of streets (Section \ref{sota.street})
			 , the reconstruction of street network (Section \ref{sota.street_network}) 
			 , the reconstruction of urban vegetation (Section \ref{sota.vegetation}) and 
			 the reconstruction of urban objects (Section \ref{sota.urban_feature}).

		\subsubsection{Multi scale} 
			The spatial extend of a city model is large (typically in the order of the $100$ to $10 000  \kilo\squaremetre$ ), yet many important part of the design are small (e.g. a curb is around $0.1 \metre$ high but strictly defines radically different space usages: pedestrian vs vehicle). 
			
		\subsubsection{Automation}
			Manual modelling as been the tool of choice for a long time but can only be applied to small parts.
			Therefore city reconstruction must use automatic or interactive tools.
			Yet the more automatic a process is, the more it relies on data quality, which is particularly problematic in an environment where aerial data is of reduced use and land data is heavily occluded.
		
		\subsubsection{Cluttering}
			Urban environment is so dense and cluttered that usually the data is only partial (e.g. a tree hides a part of a building facade).
			This is illustrated in figure \ref{sota.fig.tree_occluding}.
			\myimageHL{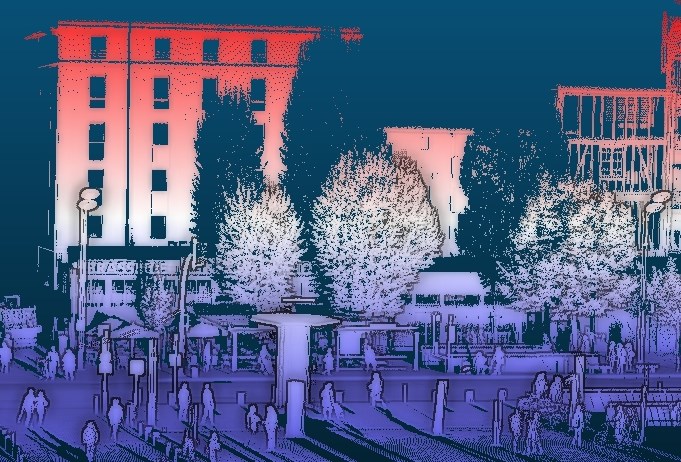}{On this street Lidar point cloud, trees are clearly masking building facades, creating occlusion.}{sota.fig.tree_occluding}{1}
	  
		\subsubsection{Many object categories}
			Cities contains many objects ("object" being used in a wide acceptance) that forms complex patterns of relations.
			For instance streets markings follow complex rules that enforce the highway code.
	\subsection{City reconstruction/modelling}
		\label{sota.intro.city_reconstruction}
		
		\subsubsection{City reconstruction}
		Several research communities have been interested in city reconstruction. It has proved to be a challenging and highly interdisciplinary set of problems, with many major practical applications.
		
		One could think the upper bound of city reconstruction problem is to have partial models of how an existing city looks as well as how it works, sometime trough time.
		
		However understanding the functioning of a city is out of our scope and this state of the art focus on reconstructing its physical components.
		
		The expression "city reconstruction" and "city modelling" are used alike to designate our problem by different research communities.
		Both convey the same idea of a partial view of reality, along with some knowledge about its structure (reconstruction) and/or behaviour (modelling).
		More practically, the model is designed to be browsed through digital imaging (which according to~\citep{Ramilo2005}, is more efficient that a real life scaled model).
		To be precise usually the goal is not exactly city reconstruction but more the reconstruction of a specific urban space with some of its structuring properties and the abstraction of key characteristics.
		
		\subsubsection{A link between modelling and reconstruction}
		The term "modelling" seems to be more used in the Computer Graphic community, while the term reconstruction seems to be more employed in the Photogrammetry and Remote Sensing community. 
		
		We use both because there is a point of convergence. 
		
		Trying to reconstruct a city and its components always involves an implicit model (or hypothesis) determined by the choices of algorithms and constraints (e.g. most of the time building are implicitly considered like locally planar blocks, mostly vertical). We observe a trend in reconstruction to use more abstract knowledge, like semantic consideration (e.g trying to reconstruct buildings parts with traditional stereo reconstruction as well as primitive fitting~\citep{Lafarge2010}).
		
		In the same time, while modelling is not exclusively dedicated to depict real world objects, it is still possible to use modelling methods to get a model as close to the reality as possible.
		It has traditionally been done with human feedback (a 3D artist uses pictures of an object and dedicated software to draw it in 3D), but a recent trend tries to do it automatically: the inverse procedural modelling paradigm (See Sec. \vref{sota.approaches.procedural_modelling}) . 
		
		The consequence is very important for this work.
		Indeed most modelling methods have the potential to be used in reconstruction process via the inverse procedural modelling paradigm, and thus we include these modelling methods in this state of the art.

	\subsubsection{Scope}
		
		In this work we focus on the reconstruction of the morphological characteristics of a city, because a full city modelling would also require to model the social and economical phenomenon, which is way out of our scope.
		
		\paragraph{Geometrical models}
			we focus on the methods to obtain models of cities. We limit the possible usages and data type to the most common.
			For instance we do not describe audio feature, even if it is an essential piece for realism (that can also limit the need for visual details~\citep{Mastoropoulou2005}).
			We focus on geometrical models.
		
		\paragraph{Not only buildings}
			Most of the works in city reconstruction have been focused on buildings reconstruction ~\citep{Musialski2012a,Klavdianos2013}.
			Yet a city is far from being only an aggregate of buildings.
			Paris, one of the densest city in the world, is a good example.
			About ~70\% of the surface is \emph{not} occupied by buildings.
			The streets occupy ~40\% and the places ~5\%.
			
			We can explain this by the fact that a city is by definition a place of complex social interactions, therefore a need for a common medium is essential and must exist: the road network and the streets.
			
			\myimageHL{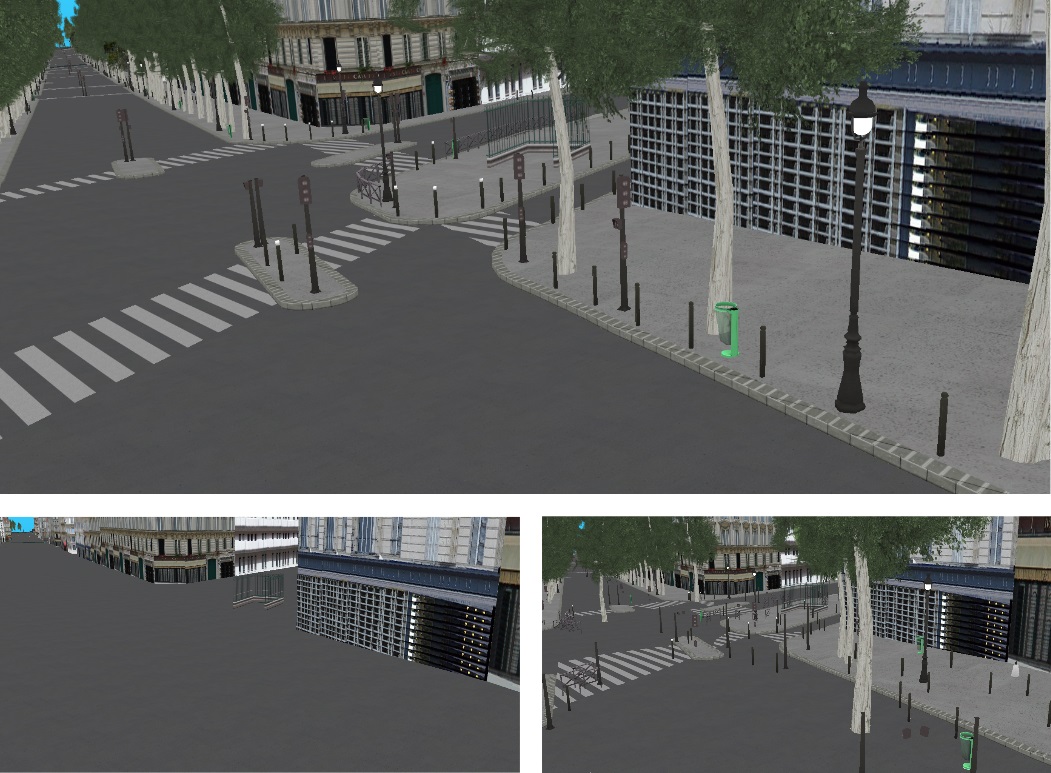}{Synthetic 3D city model. When urban features are coherent (top), the model is a great deal more realistic than without any objects (bottom left), or even with the same amount of objects but un-organised (bottom right).}{intro.fig.w_wo_objects}{1}
			
			Moreover, a crude perceptive example (Figure \ref{intro.fig.w_wo_objects}) shows the importance of de road network, street, vegetation, urban objects.
			
			A real street view of Toulouse city (Fig \vref{intro.fig.street_objects}) with urban objects highlighted shows the diversity and importance of street objects.
			 \myimage{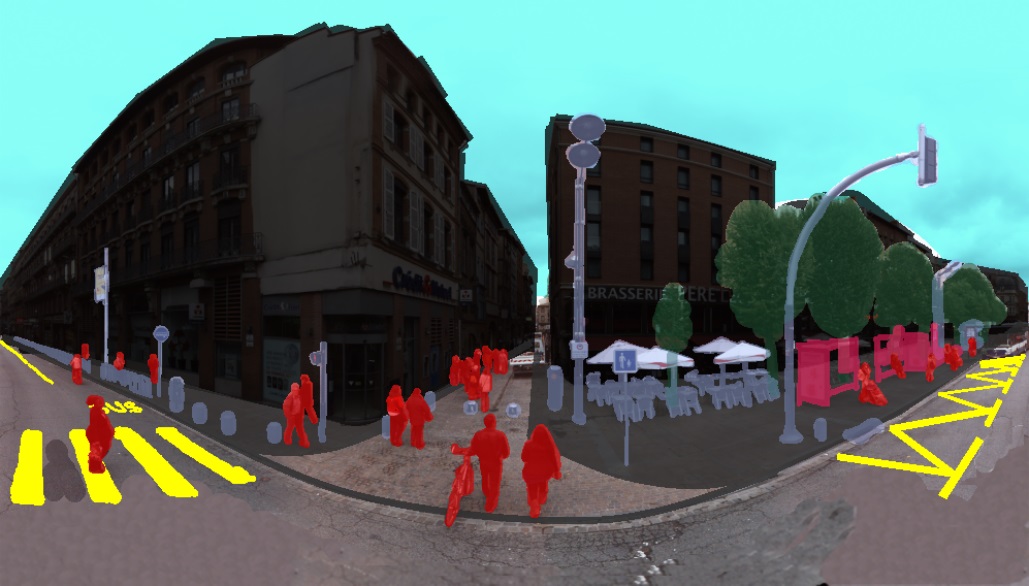}{importance objets}{intro.fig.street_objects}
			 Almost all the applications of city modelling (See Figure \vref{sota.fig.3d_model_usage}) 
			  benefit as well from such additions.
			Another clue of the importance of non-building elements for city modelling can be given by analysis of City GML~\citep{Kolbe2005}, the leading current standard to represent city.
			Building is only one City GML module among a dozen other (transportation, vegetation, urban furniture ...).
		
		\paragraph{No transport simulation}
			Crowd and traffic simulation is out of the scope of this work. However, such simulation necessitate reconstruction of specific data which we will briefly cover. 
			Reader can refer to the recent state of the art of~\citep{Duives2013} for more details about traffic simulation.

	\subsection{Plan}
	
		Reconstruction always necessitates data representing the city, we first analyse which data are available (\ref{sota.input_data}).
		City is composed of many components, and many methods from different research community try to reconstruct them.
		We first propose a trans-components classification of reconstruction approaches (\ref{sota.approaches}). 
		We then dress independent state of the art for each category of city components being reconstructed, such as street reconstruction (Section \vref{sota.street}),  street network reconstruction (\vref{sota.street_network}), urban features reconstruction/modelling (vegetation (\vref{sota.vegetation}), and urban objects reconstruction/modelling \vref{sota.urban_feature}). 
		This simple ordering is necessary due to the wide differences between methods.
		For each of this categories, we propose several ways to classify the state of the art methods, to allow a multi-level understanding of the field.
 
		We conclude this work by giving perspectives about the evolution of city reconstruction.


\section{Input data}
	  \label{sota.input_data}
	We analyse what data are available for city components reconstruction.
	These data can be ordered from less structured (Lidar data, vector data) to more structured (image data, raster data), and from less abstract (Lidar data, image data) to more abstract (raster data, vector data).
	(See Figure \ref{sota.fig.structured_vs_abstract})
	 
	\myimageInline{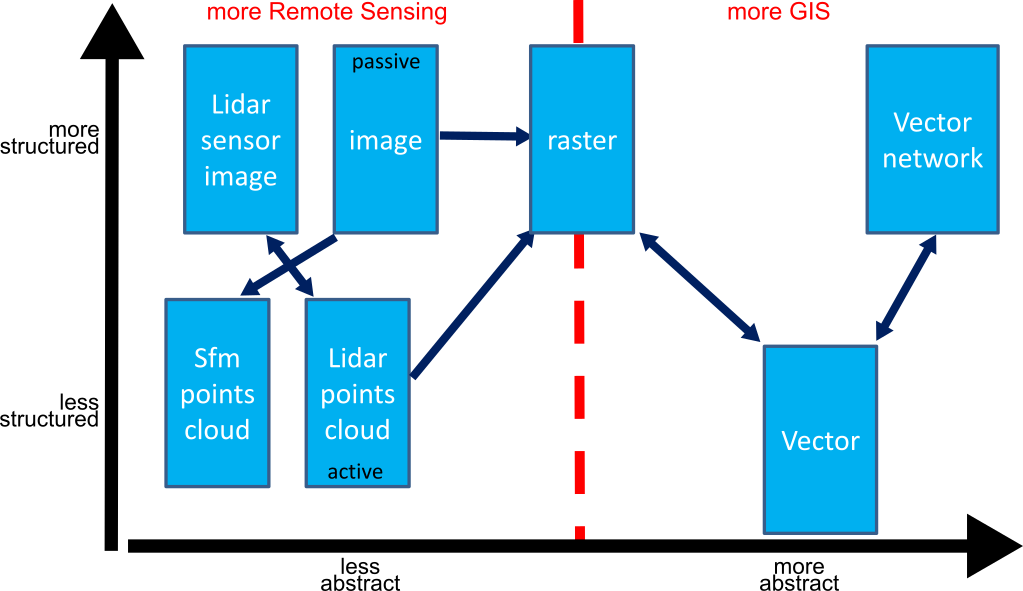}{Available data types for city components reconstruction, ordered from less to more abstract, and from less to more structured.}{sota.fig.structured_vs_abstract} 
	  
	Please note that this classification is based on common usage rather than on strict mathematical differences.
	From a mathematical point of view, Lidar, image and raster data are of same nature (2D lattices, i.e. regularly sampled values), and the definition of vectors data is vague (parametrized shapes with semantic, the types of shape and parameters varying).
	
	Moreover the boundaries may be fuzzy.
	For example it is possible to create an image from a Lidar pointcloud (sensor view, See Fig. \vref{sota.idata.lidar_image}), and a pointcloud from multi-images (Structure From Motion, SFM, see \cite{Carrivick2016} for a recent book about sfm and geoscience ). 
	Similarly, a conservative two-way conversion between raster and vector data is possible under certain assumptions (2.5D). 
	 
	In the next sections we introduce each of this available data.
	Each section is illustrated by real data for a street of Toulouse city (France), from mobile mapping (\cite{Paparoditis2012}), or from aerial images (French mapping agency, IGN).
	
	\subsection{Lidar data} 
		\myimageFullPageWidth{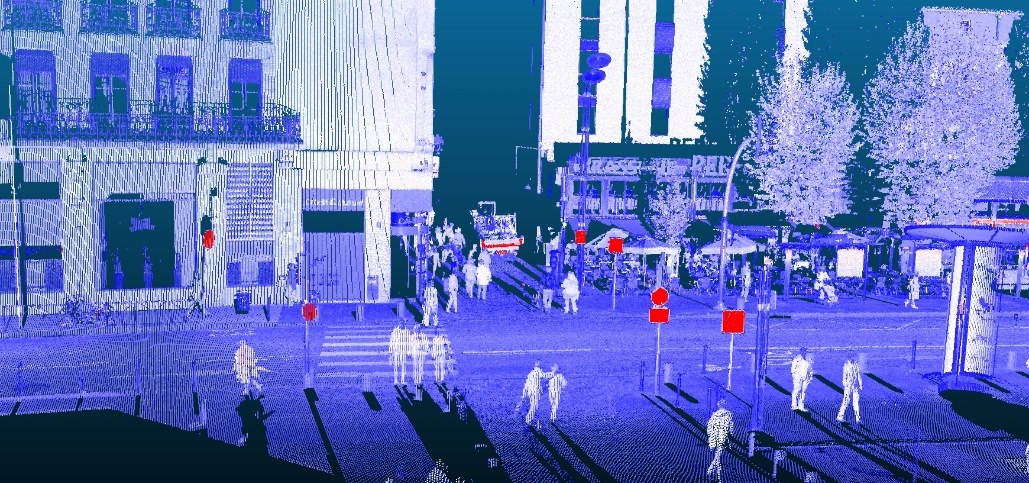}{Street Lidar Point cloud ( intensity tone from blue to white to red).}{sota.fig.lidar}
	 	
		\subsubsection{Intro} 
			Light Detection And Ranging (LIDAR, noted "Lidar" for readability) data are becoming more and more available to the point of being common. Their interest is partly due to the complementary nature they offer to images, making them extremely popular in urban reconstruction. In particular, they offer easy access to 3D coordinates.
			A Lidar device uses active sensing, and can be fix or embedded on mobile objects (plan, drone, vehicles, train, etc.).
			Figure \ref{sota.fig.lidar} illustrates a point cloud from a terrestrial Lidar.
	 	
		\subsubsection{Principle}  
			The Lidar principle is simple and very similar to a Laser measuring tape. The device emits a short light impulsion (i.e. active sensing) from a know position in a known direction at a precisely known time.
			This light signal flies for some time, hits an obstacle and is partially reflected backward to the device. The device receives this reflected signal. Then it analyses the time of flight, and given the speed of light in air, it can compute the distance from the device to the obstacle. This gives the precise 3D position of the obstacle, hence a 3D point.
	
			The magnitude (i.e amplitude of signal) of the return impulse is also extracted, quantifying the ability of the obstacle to efficiently reflect light (at the Lidar frequency, for a given input angle). Intuitively, a street furniture of polished metal will reflect much more light than a rugged stone wall.
			 	
		\subsubsection{Data volume} 
			Such sensing is made at high frequency (0.1 to 1 $\mega pts \per \second$), making the data volume huge and barely tractable in practice. 
			Yet, even at several millions of point per second, we are short of a typical HD video-film acquisition data rate (1200*1800 pixels, 25 times per second). However, the data volume is much more difficult to manage with Lidar data than image data (See \cite{Cura2016b} for more details about point cloud management). 
			
			One has to remember than photography have been invented two centuries ago, and that digital imaging has been researched for several decades.
			In opposition dense 3D point clouds and Lidar processing are much newer.
			The industry still lacks standard formats, powerful viewers and editors, and mature compressions (for example : ~\citep{Mongus2011}. The link to the compressive sensing theory~\citep{Baraniuk2011} does not seem to have attracted much interest either). 
			
			All in all, the main issue with Lidar data is no more its volume, but the lack of management framework as a whole (See \cite{Cura2016b}).
			
		\subsubsection{Details and facts} 	
			
			Lidar can be airborne (several points per square meter, precision of $0.1$ to $1 \metre$) or ground based (for stationary station: precision less than $1 \milli\metre$, for vehicle: precision around $0.1$ to $1 \centi\meter$).
			
			More sophisticated methods allow to acquire and store the full waveform of the return signal, which can be used to extract multiple points per waveform (e.g. one point for the forest canopy, one point for the forest mid level and one point for the ground)~\citep{Mallet2010}.
			Other recent technologies propose multi-spectrums Lidar~\citep{Hakala2012,Wallace2012}, but these remain in laboratories for the moment.
			
			It is important to note that LIDAR data is an accurate and sparse sampling of 3D objects by nature (e.g. the size of error is small compared to the distance between points). 
			As such, their sparse and 3D nature is complementary to the high density and 2D nature of images. 
			
			\myimageHL{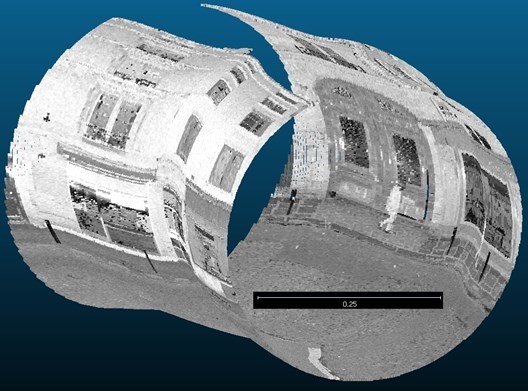}{Lidar points can be seen as images (sensor image) as the acquisition process is regular. }{sota.idata.lidar_image}{1}
			
			It is possible to create an image from the Lidar device point of view (sensor image, see Figure \ref{sota.idata.lidar_image}), because the device physically acquires points following a very regular pattern (lines, grids).
			Such an image is equivalent (dual) to the traditional 3D point cloud, and can be used as a 3D depth map. 
			However it is not commonly used because working in image space disables the possibility to use classic Euclidian distances.
			There is no straight relation between 3D distance and pixel distance (2 points separated by 6 meters in 3D world may be separated by any number of pixels).

	\subsection{Images}
		\label{sota.input_data.images}
		\myimageFullPageWidth{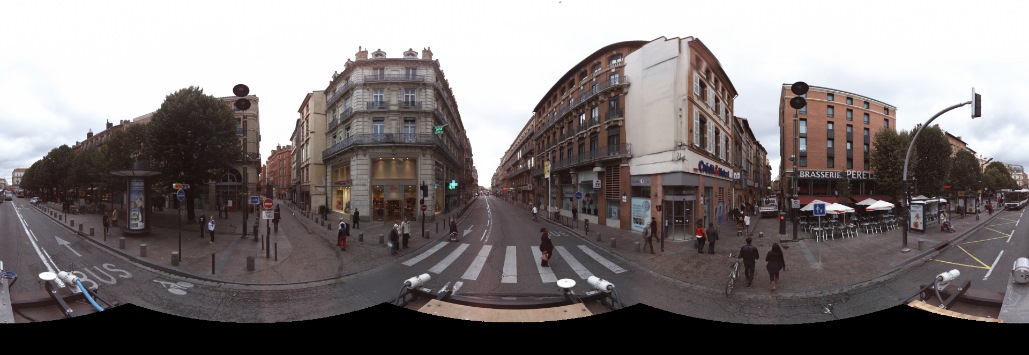}{Street View (360 degrees panoramic) .}{sota.fig.mobile_mapping}{0.95}
			
		\subsubsection{Introduction}	 	
			Images are very common data for urban reconstruction, partly because they are widely available and have been used in computer science for a long time, and also because they are so similar to how we view our environment.
			Images can be street view images (See Figure \ref{sota.fig.mobile_mapping}) or aerial images (See Figure \ref{sota.fig.aerial_view}). 
			Efficient image processing is made possible by the very regular nature of image data (in particular, pixel neighbourhood is known) and dedicated powerful graphics hardware (Graphics Cards with dedicated in silico parallel processing pipelines).

		\subsubsection{Principle}
			Image sensing is an approximation of the complex nature of the light signal emitted/reflected by an object at a given time. A camera is only a receiving sensors, making it a passive method. 
			
			The camera has an array of photo-sensible sensors. Each sensor counts the number of photons arriving during a given time, thus gives an average intensity of light signal over a short time. Matrices of coloured filters allow acquiring the intensity of different parts of the light spectrum (e.g. (Red,Green,Blue) colours).
			
		\subsubsection{Aerial image}  
			\myimageHL{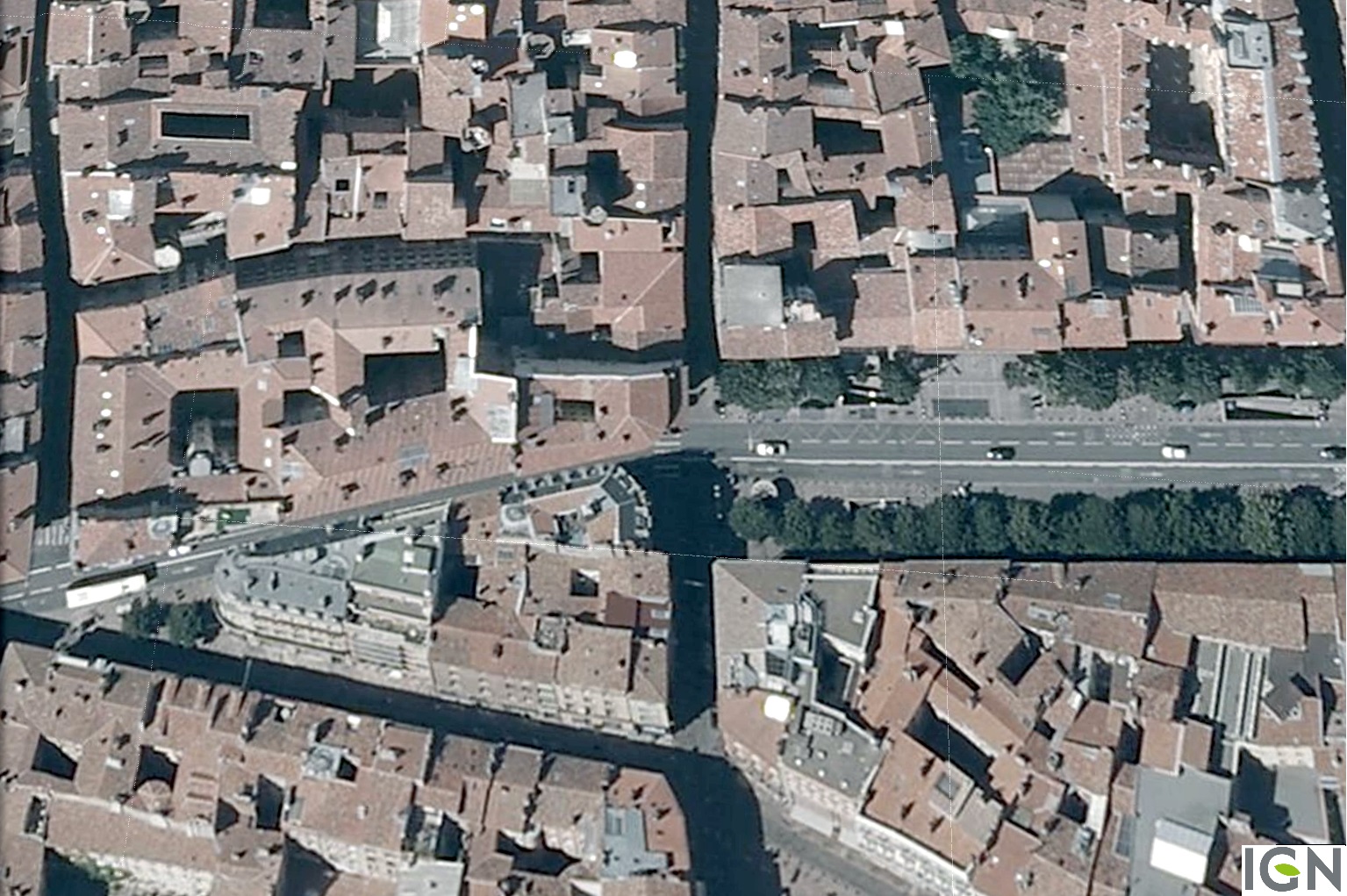}{Aerial image data.}{sota.fig.aerial_view}{1}
		
			We differentiate aerial images (See Figure \ref{sota.fig.aerial_view}) from street images as the cameras are usually significantly different.
			
			Aerial images (satellite and viewed from planes) have extensive geographic coverage because of the near uniform acquisition process and many satellites available. In urban context, they tend to have massive occlusion because street canyons are occluding parts of the city.
			Several passes with different acquisition angles and directions can help reduce this problem~\citep{Garcia-Dorado2013}. 
			
			Radiometric quality is generally high and distortion low. These images are precisely geolocated (i.e. we precisely know from where and in which direction the image was taken in relation to known ground features). 
			
			Other spectrum than human-visible colours are often available and give precious information (e.g. near infra-red for tree detection). Pixel width is typically between 0.1 and 1 meter.
		
		\subsubsection{Street image}
			
			Street views (\ref{sota.fig.mobile_mapping}) are usually taken from the ground by a person or a dedicated vehicle circulating the streets.
			These images allow seeing in great detail buildings façades and streets. 
			
			Geolocation of images is done trough GPS and inertia sensors, but a centimetric registration of those images is still an open problem.
			
			It may be hard to coherently use multiple images because of the level of change and imprecision. 
			
			Spatial coherence is difficult to obtain because of the registration challenge, temporal coherence because significant parts of the images may be occluded by moving objects, and radiometric coherence (colours) because the lighting conditions may change very quickly (e.g. moving from shadow to direct sun light).
			
			The pixel size varies with the depth of the image but can be estimated from $1$ to $10 \centi\metre$ average, and the data volume is very sizeable (thousands of images per hour).

			It is common to use multiple images to create sparse point clouds using a method called "structure from motion" (SfM). Such point cloud can then be densified through dense matching.
			However such point-clouds are very different from those obtained by Lidar.
			Due to instrinsic 3D reconstruction ambiguities, errors and noise are typically higher, and point sets are sparser in uniform area (e.g. a white texture-less wall).
			This differences partially explains why many methods are specific to Lidar pointcloud or SfM pointcloud.
			\citep[Sec. 2.2]{Musialski2012} give an introduction to SFM.

	\subsection{Raster data}
		
		\myimageHL{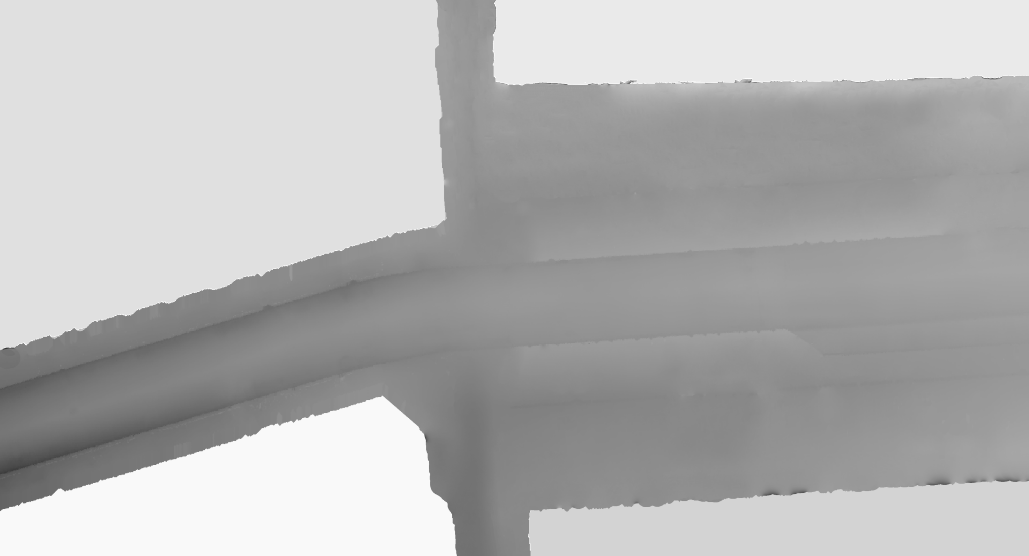}{Raster Digital Terrain Model (DTM) data.}{sota.fig.dtm}{1}
			
		\subsubsection{Introduction} 
			In Geographical Information Science (GIS), a raster dataset is a regularly sampled 2D distribution (i.e an image) draped over a portion of ground ( i.e viewed from above). 
			This image can contain colors channel (RGB), but also any sampled field values.
			
			The information contained can then simply be a visual texture (an ortho-photography, making it close to images (Section \ref{sota.input_data.images}).
			But it can also be more abstract, giving for example some geometric information (the estimated height of roofs in a given area), or semantic information (the probability that the space the  pixels cover are made of vegetation), or even statistical (an estimation of the average income distribution over a city).
			The Figure \ref{sota.fig.dtm} shows a raster representing the ground elevation.

		\subsubsection{Level Of Details (LOD)}
			Rasters can be very large images (covering large ground with many small pixels). Thus they typically require a Level Of Details (LOD ) approach. 
			A raster can be tiled (regularly cut into smaller pieces) to access only a part of interest, and/or used at varied resolution (e.g. pyramid representation of the JPEG 2000 standard).
			Note that point cloud can also use LOD approach (See \cite{Cura2016b}).

		\subsubsection{Details}
	
			Rasters are prone to quantization errors (intuitively, it is hard to represent a curve with rectangular pixels). 
			They also have an obvious limitation: they can represent a  2.5D surface but not all 3D form or volume.
			Full 3D volume can be obtained with voxel grids, but these does not seem to be as used as plain rasters.
				 
			It is important to note that raster can be more semantically abstract than images and Lidar because they can sample any distribution. Thus the values can represent other things than direct sensing data (e.g. a land use map).

	\subsection{Vector data}
	
		\myimageHL{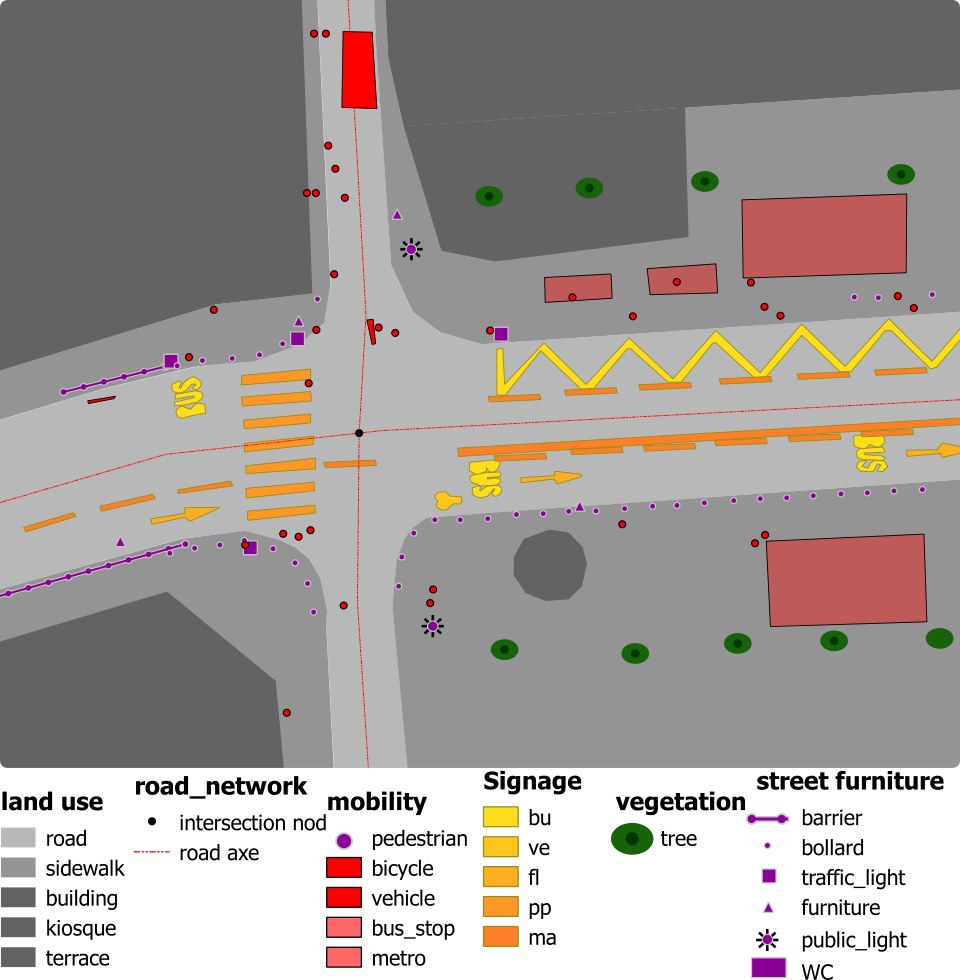}{Vector data: various vector types forming a map.}{sota.fig.vector_street}{1} 
			
		\subsubsection{Intro}
		
			Vector Data are classical data in map making (See Figure \ref{sota.fig.vector_street} which illustrates a number of vector data forming a map).
			Intuitively, vector data are arbitrary (mostly 2D) simple parametrized shapes with attributes, and often associated visualisation rules.
			Usually the shapes are limited to points, poly-lines and polygons with holes. More generic forms like curve (arc of circles,Bezier curves, splines) , or 3D primitives (meshes, triangulated networks) are less common.\\
			Attributes are values attached to a shape. For instance a tree could be represented as a polygon for the boundary of its trunk at ground height, along with the attribute ''tree\_species'' (e.g. "Platanus") and ''height\_of\_the\_tree'' (e.g. "12.4").
			Vectors are closely associated with maps, therefore they are often used in complex visualisation rules. 
			For instance a point with a text attribute can be visualised as a label (text on map).

		\subsubsection{Abstract data}
			
			Vector data are usually more abstract than the other data by nature and by usage.
			By nature because both images and lidar point clouds can be represented losslessly with vectors.
			However vector data is irregular by nature (no information on neighbourhood). As a consequence representing images as grid of rectangle vectors is of reduced interest.
			
			Vector data are also traditionally more abstract by usage. For instance a 2D polygon would be very classically used to represent a building footprint. In this case this polygon is already a simplified model of the building. Given the height, we could extrude the footprint to create a simple building volume.

		\subsubsection{Obtaining vectors}
		
			Vectors are not a direct result of sensing, but an interpretation of reality. 
			Such interpretation can be automatic or done by human.
			Thus vector data can be obtained by an analysis of direct sensing data, which is a part of the objectives of the remote sensing research community (e.g road extraction~\citep{BarHillel2012}). 
			
			Vectors can also be man-drawn (using aerial images in background), produced by field survey (involving a positional device to map the objects), or extracted from pre-existing maps (vectorisation).

			Vectors are not regular by nature (no information about neighbourhood), however using the attributes one can create a so called "topological model", which allows to create a graph structure over vectors. 
			For instance a road axis network is composed of road axis (vector) with topological information (axis $i$ and $j$ are connected at node $n$).
			
			A typical example would be that for each vector an attributes gives its connected vectors with some orientation information. This data structure is harder to manage and to use but allows minimizing data duplication. Using such a data structure enable different applications like traffic simulation, or advanced spatial-relationship analysis.


\section{Approaches for reconstruction / modelling}
\label{sota.approaches}
In this section, we first propose a transverse classification of methods reconstructing/modelling the different urban aspects.
Then, we give pointers for procedural modelling, grammar, and inverse procedural modelling.

\subsection{Transverse reconstruction method classification} 
	
	\subsubsection{Reconstruction strategies ordered by model importance}
	In the rest of this chapter, we consider a short state of the art for each aspect of urban modelling/reconstruction (building reconstruction (Sec. \ref{sota.building}), street reconstruction (Sec. \ref{sota.street}), street network reconstruction (Sec. \ref{sota.street_network}), urban features reconstruction/modelling (vegetation (Sec. \ref{sota.vegetation}), and urban objects reconstruction/modelling (Sec. \ref{sota.urban_feature}).
	
	These aspects are very different in nature, type of data used, and type of results. 
	Therefore we propose several classifications for the methods of each aspects.
	Each classification is intended as a way to compare methods.
	
	All methods deal with the reconstruction of one aspect of urban model, we propose a transverse classification of these methods.
	We choose to classify reconstruction strategies by the role the model play in the reconstruction method. 
	At one end of the spectrum, the strategy of direct reconstruction from sensing data (e.g. triangulate a point cloud for instance).
	In this strategy, the model has a very small role, as it is mostly implicit.
	
	At the other end of the spectrum, the strategy of catalogue matching. In this case the reconstruction strategy is to identify which model represents best the data, therefore, the model play a very large role.
	
	This classification is illustrated in Figure \vref{sota.fig:transverse_reconstruction}.
	
		\paragraph{Data-driven reconstruction}
		Some methods reconstruct directly from sensing data (low level reconstruction, data-driven), for instance reconstructing the ground surface, the building approximate geometry, the road network from image, etc.
		Similarly, points or pixels classification can be seen as low level reconstruction.
		These methods have the advantage to rely on an implicit model which may be very generic.
		Yet, the sensing data is often sparse and of relative low quality considering the scale of the considered objects.
		Moreover, low level reconstruction methods seems to be ill adapted to output structured/complex results (for instance a facade organisation, a hierarchical road network, a graph of parts of a man made object, etc.).
		
		\paragraph{Model-driven reconstruction}
		A way to simplify a problem too wide is to add constraints and knowledge about it. 
		Some of the methods therefore add strong hypothesis about the object to reconstruct, typically exploiting prior knowledge (road slope and turning radius is constrained by civil engineering rules, trees tends to grow to maximise exposition to sun light, etc.), and hypothesis of symmetries. 
		
		These prior knowledge are then expressed as strong models (Template, pattern, etc), and the reconstruction is much more model-driven (top-down).
		For instance when reconstructing road markings of pedestrian crossing, we can use the hypothesis that each strip is a rectangle, and that related strips are parallel with a regular spacing.
		
		\paragraph{Procedural modelling}
		However template and pattern become difficult to use when the reconstructed objects follow complex patterns and/or hierarchical patterns.
		In this case, procedural model offers a powerful and adaptable way to construct such results (for instance, expressing a tree procedurally).
		
		When reconstructed objects have important and structuring relationship, a grammar is a good tool to formalise these while keeping a strong modelling power (for instance, using a facade grammar, shutters would necessarily be created and linked to a window).
		Moreover grammar are very hierarchical by nature, which suits well a number of aspects of urban reconstruction, as both natural and man-made object express .
		
		\paragraph{Inverse Procedural Modelling}
		Procedural modelling and grammar modelling have great modelling power, but are hard to use in reconstruction.
		Indeed, they can be used to create a model, but are hard to adapt to model something in particular.
		In this case the paradigm of Inverse Procedural Modelling is necessary, that is given a model and observations of the object to be reconstructed, what are the parameters and rules of the model that best suits the observations (for instance, given a pedestrian crossing detection, what is its orientation, width, number of bands, etc.).
		The number of parameters to consider is extremely large, and this, in addition to sparse and noisy observations, may lead to an intractable problem.

		\paragraph{Catalogue matching}
		In some case, the objects to reconstruct are very well known and may have very little variations. 
		Thus, we can adopt a catalogue matching strategy. 
		Instead of reconstructing an object, we use observations of the object to find the model that is the best match in a large model database.
		For instance, using a streetview we detect a urban furniture.
		The image is matched to a database of 3D model of street furniture.
		The best model is then scaled and oriented.
		Please note that in this case, the model almost totally determine the result. 
		This allows to decompose the reconstruction problem: First find which street object is where, possibly determining some of its properties, such as its orientation. 
		Then, find or generate a similar 3D model and populate the reconstructed street with it.

	\subsubsection{Additional considerations for reconstruction strategies} 
	
		\paragraph{Interaction}
		Independently of the strategy used to reconstruct object, a user interaction is often necessary.
		This is especially the case when input data can not be really trusted, or when the reconstruction method strongly relies on model (procedural modelling for instance).
		Controlling grammar is difficult and dedicated methods may have to be tailored (for instance, using brush to describe the different parts of a city; or a street network may be generated basing its morphology onto the surroundings).
		
		\paragraph{Updating database and fusion}
		Most of the methods we presented are straightforward modelling/reconstruction methods working on sensing data. 
		However, for real life application (especially street network reconstruction), one may use not only sensing data, but also a previous coarser results.
		For instance an incomplete road network is completed with road extracted from sensing data.
		These methods are still about reconstruction, but they may also contains supplementary parts such as data fusion, data qualification, etc.

\subsection{Procedural modelling and grammar}
\label{sota.approaches.procedural_modelling}

In the procedural modelling paradigm, a model is not defined by a set a parameters, but by a set of rules that can be combined, for instance in a grammar, to model complex objects.
This type of modelling has a very high descriptive power, which can be hinted by the fact that grammars are at the very basis of how we express ourselves, and at the veryc ore of computer science.

We recommend the read of the seminal article of \cite{Parish2001, Muller2006} for an introduction to shape grammar (The Figure \ref{sota.fig:shape_grammar} is extracted from \cite{Parish2001}).
\myimageHL{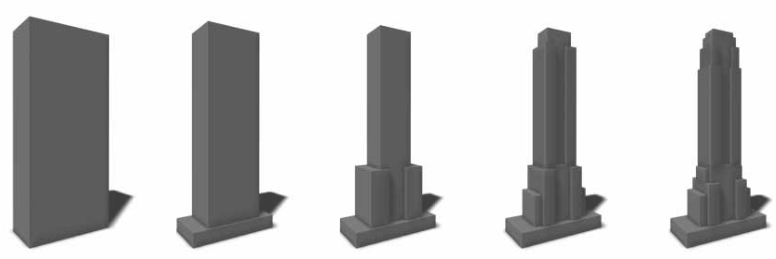}{An example of usage of shape grammar from the seminal article of \cite{Parish2001}.}{sota.fig:shape_grammar}{1}

\subsection{Inverse procedural modelling}
\label{sota.approaches.inverse_procedural_modelling}

Inverse procedural modelling is the paradigm where a procedural model is fitted to observations.
It is important to note that we do not only look for the parameters of the model, but also for the rules used in the model (i.e. the number of parameters is not fixed).
For instance in the case of a facade grammar, we do not only look for the number of floors, but for the rules that will be used to generate these floors (for instance, create window with balcony and shutter).


\section{Buildings and façades}
\label{sota.building}
Building reconstruction has received much attention in the past decade.
Thus, methods have focused on diverse parts of buildings reconstruction (facade reconstruction, roof reconstruction, indoor reconstruction, etc.).

Different types of building may also be reconstructed using different methods (Manhattan  /Atlanta /Planar-hinged building type~\citep{Garcia-Dorado2013} or suburban house~\citep{Lin2013}).
Some methods focus on large scale solution, efficient visualization, Level Of Detail feature, etc.

The methods used are so diverse that the author of the recent state of the art~\citep{Musialski2012} have chosen a straight order by goal and data input.
\cite{Klavdianos2013} also establish a building reconstruction state of the art.

\myimageHL{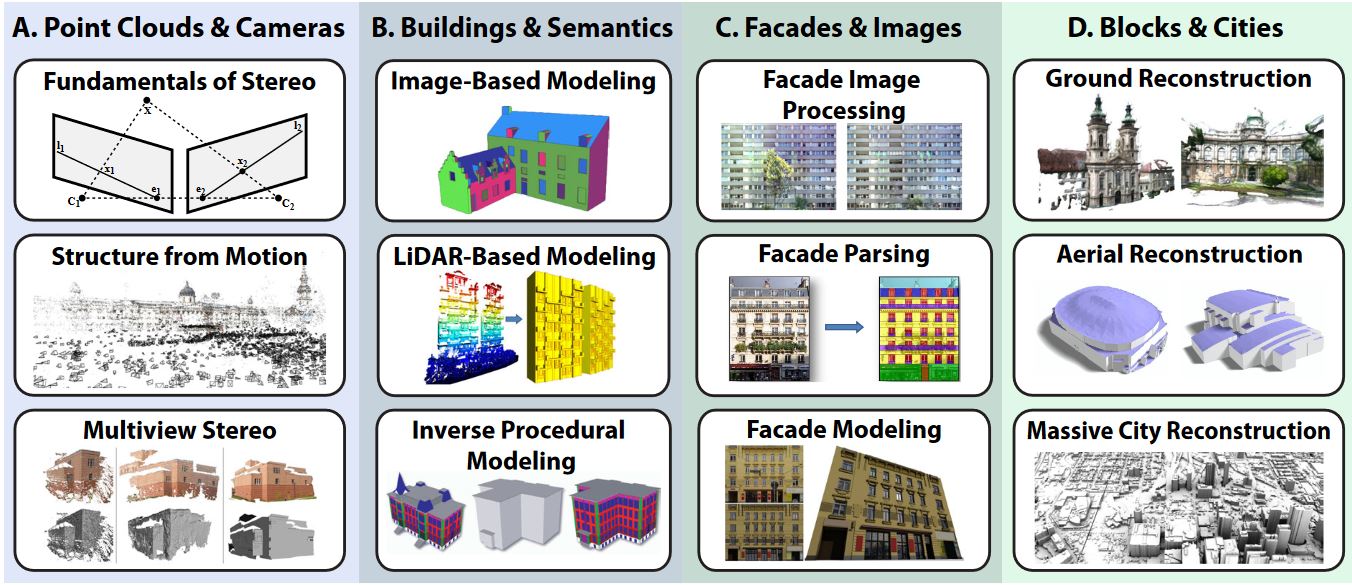}{Illustration of \cite{Musialski2012} illustrating building reconstruction.}{sota.building.musialski}{1}  

We refer to the Figure \ref{sota.building.musialski} extracted from \citep{Musialski2012} for a quick overview of different approaches for building reconstruction.

In this work we chose to not develop this topic, as it is covered by recent states of the arts articles for building reconstruction. 
We note that many of the strategies explained in these articles could be used for the reconstruction of other objects.
We also feel that the building reconstruction community has pioneered many advanced articles about shape grammar and inverse procedural modelling.
 

\section{Street}
\label{sota.street}

\myimageHL{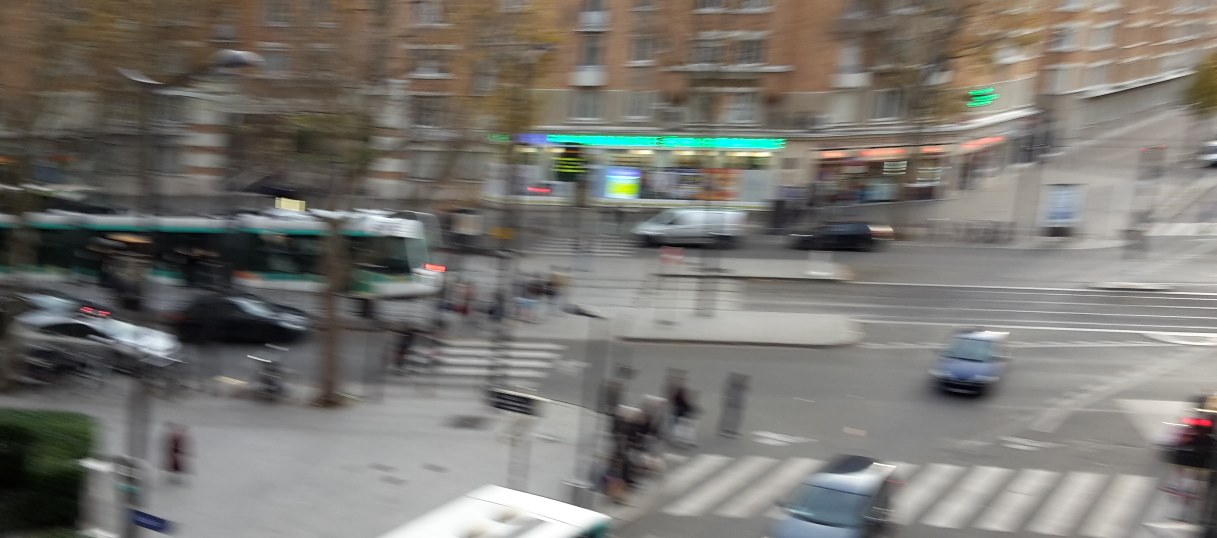}{One of Paris street.}{sota.streets.didot}{1}  
 
	\subsection{Introduction to street reconstruction} 
	\subsubsection{Challenges for street reconstruction}
	Streets are essential components of a city model. As the medium pervading all other structures and objects they are complex. 
	First the geometric nature of streets is specific, detailed and not normalized. \\
 	Second, a street is a complex arrangement of objects that are inter-related and have their own structure. For instance a pedestrian crossing is located in relation to traffic light and is a structured composition of markings bands.\\
	Third, streets are objects that are strongly defined by the uses the inhabitants of the city make for it, in particular regarding their displacement.
	
	Thus, a street organisation is partly guided by these functions, and as such, street reconstruction should provide an ouput compatible with this functions. 
	
	\paragraph{Streets are complex, even for human}
	Streets are so familiar places that we specialise very early in using them during childhood.
	However one can remember the complexity of the task when travelling in another country.
	There, every aspect of a street can be different.
	
	Children have to be taught a long time where and when to walk, not speaking about driving rules, or using the public transportation system, which are even more complex tasks.
	In this spirit, people with even a light intellectual or physical disability may have significant trouble navigating the public transportation system, which is based on streets.
	
	\paragraph{Street for traffic}
	An essential function of street is multi modal navigation (vehicle, public transport, bikes and pedestrian). Such navigation uses network level features (Section \ref{sota.street_network}) which have great impact at the street level organisation. For instance the sole purpose of street markings is to support traffics. 
		Being on the ground they are prone to occlusion and wear, but there use is strictly regulated (e.g. France reference document is 65 pages long, (French Ministry, 2012)).
		
	Streets are used for several transportation methods (pedestrian, bike, public transport, vehicle) that are mixed (e.g. a pedestrian crossing is shared between pedestrians and vehicles).These methods shapes in turn the streets, which add to the complexity of it, and increases the difficulty to reconstruct the street.
	
	\paragraph{Streets are organised}
	Streets are challenging to model because they form a partially organized structure (typically organised relatively to the central axis), yet are much less locally regular than a building (in particular, the relations are more generalised, like intrinsic partial symmetry vs extrinsic, see~\citep{Mitra2012}, ch. 7). 
	Also, the street components have strong relations between them, which makes difficult to model a small area at time (in opposition to building which can reasonably be defined as dissociated from the other close buildings). 
	For example, reconstructing a pedestrian crossing usually implies there would be another in the next hundreds of meters.
	Some street features may follow a partial symmetry (bollards for instance), a pattern (pedestrian crossing markings bands), or be organised in inter-related hierarchy (lane markings and traffic light) .

	\paragraph{Streets are hard to sense}
	Lastly the data collection is difficult. 
	Aerial sensing may be impeded by buildings, and the street geometry and features make it difficult to avoid large occlusions due to traffic, people, trees...
	In opposition to building, whose main feature (door, windows, etc) are large (1\metre), street are partially organised by kerb (separator between road and sidewalk) which are much smaller (0.1\metre).
	The necessary geometric precision is even greater when considering slope and water drainage.

	\paragraph{Related work}
	
	There are been remarkably few works on reconstructing streets, even if streets contribution to a city model is evident.
	We can conjecture that this is partly due to the fact that data to the required precision (less than $0.1 \metre$ for a basic curb) has been only recently available in urban environment.
	Also, the geometric nature (streets are not necessary blocky and have irregular shapes), the diverse and complex arrangement of objects (markings, signs, furniture), and the dependency on many research fields (object detection \& segmentation, pattern recognition, paving and texturing, intrinsic symmetry detection) makes the problem challenging.

	However the simulation industry has used dedicated data models (for instance, RoadXML\footnote{\url{www.road-xml.org}} or OpenDrive\footnote{\url{www.opendrive.org}}), which characteristically include network aspect, surface material and 3D representation, along with road objects and road related objects.
	Powerful specialised softwares allow to design intersections in all their aspects (lane size \& position, traffic regulation depending on the traffic throughput, regulatory material) in the construction and CAD field. 
	These software are not included in this state of the art, as we were not able to test them.
	They also seem to be more designed oriented than reconstruction oriented.

	Street are also part of a street network, this constraint needs to be enforced at all time, but can provide precious information (e.g. traffic direction(s), thus orientation of traffic signs, etc.). Therefore some methods for street network modelling can also be partially applied to model streets.
	Even if it has been a common practice, modelling streets like the complimentary space of buildings is not sufficient~\citep{Cornelis2008} for many applications, and in many cases simply erroneous (e.g. private garden, places, parks).

	\subsection{Modelling the geometry of the street}
	One should model the detailed geometry of the street and curb (which is typically varying to separate pedestrian crossings or driveway entrances).
	This is counter-intuitively difficult. Such process cannot rely on Manhattan-like hypothesis, and must deal with the precision issue.
	
	\subsubsection{Modelling geometry using primitives}
	Some road model from road network modelling methods can be applied (See Section \vref{sota.street_network})
	For instance road geometry can be modelled as 3D clothoid~\citep{McCrae2009, Galin2010, Applegate2011, Bertails-Descoubes2012}, arc and line pieces~\citep{Wilkie2012}, polynomial model ~\citep{Hervieu2013}, B-Spline (local only)~\citep{Wedel2009}, using road profile~\citep{Despine2011}, or using brute mesh~\citep{Cabral2009}.
	
	\subsubsection{Modelling geometry using 2D/3D grids}
	
	The street / road / ground model can also be far less constrained, and simply be a 2D or 3D grid. 
	This is very similar to having a raster with a semantic label such as road/no road.
	In this case the mode is implicit (for instance you cannot say directly that the road is of width X at this place).
	
	This type of low level modelling as been especially used for autonomous vehicles (see \cite{BarHillel2012} for a state of the art on road and lane detection for autonomous vehicle.
	There are several ways to label the space as in or outside road. 
	In mobile mapping, one can use a direct approach based on the expected height profile of the road, both from Lidar (\cite{Yu2007}).
	\cite{Cornelis2008} carve space and so model more the free space than the road. 
	
	Another way to create these maps of road surface is to classify 2D rasters representing  the scene viewed from above (\cite{Fischler1981}). 
	This raster can come from various sources, such as aerial image, aerial lidar, or be the result of another process of mobile mapping data (\cite{Serna2014}).
	
	The classification process to decide if a pixel of such a raster is to be labelled as road or not is often contextual,
	in the sense that the value of this pixel may not be sufficient, but the neighbours values may also be required.
	For this reason, the classification process is often only the first step of a more complex workflow (\cite{Montoya-Zegarra2014, Boyko2011}) that will use implicit hypothesis about a road geometry.
	For instance, cite{Boyko2011} use an active contour to find limit of road, which implicitly model the road border as smooth.
	
	The fact that the road is part of a network provides another contextual information that can be leveraged (See part \ref{sota.street_network}).
	
	However reaching the required precision might be difficult. 
	In fact, even with massive terrestrial data, automatically dealing with occlusion to get a coherent street geometry is still an open problem~\citep{Hervieu2013, Serna2013}.

	\subsection{Object detection, primitive extraction}
	But street are also a subtle arrangement of related objects.
	Street objects (like vegetation (Section \ref{sota.vegetation}), street furnitures and markings (Section \ref{sota.urban_feature})
	are hard to deal with individually.
	Detection is already a hard problem, reconstruction is even harder. 
	Detecting objects in street is challenging due to variety of objects and occlusion.
	~\citep{Golovinskiy2009} use a four steps method to detect objects in street Lidar: localisation of objects, segmentation, feature extraction and classification for a small amount of objects. 
	Beside comprehensive testing and proposing several alternatives regarding classification methods, they also reach the same conclusion about the importance of relations between objects and use ad-hoc features to this end ("contextual feature").
	\citep{Zhang2010} demonstrate the possibility to perform urban segmentation based on depth map extracted from video.
	Local features are extracted from depth map (height, planarity, distance to camera), then a random forest classifier followed by a graph cut minimization methods output a labelled segmentation.
	\citep{Yu2011} focus more on segmentation with a basic classification, but their method could be used as primary step for detailed classification.
	Similarly,~\citep{Lafarge2013} automatically extract primitives (e.g. plan cylinder, torus, etc.) from point cloud obtained by SfM. Their goal is more toward mesh compression and partial holes filling, but such primitives could also be used for object segmentation. Although the core of their method is a planar-based residential house reconstruction,~\citep{Lin2013} also detect objects (mailbox, plant, road sign, streetlight, waste bin) using an adapted version of~\citep{Zhang2010}.

	Whatever the method, the number of types of object detected is small (about 10) and the error rate varies a lot depending on type of objects. 
	The research field of fa\c{c}ade reconstruction had the same type problem.
	The trend to resolve it has been toward leveraging the organisation and relations of objects (contextual information).
	
	Therefore we include in this state of the art a prospective consideration of street related object relation detection and analysis.
	
	\subsection{Relation between objects}
	At the street level it is possible to leverage the pattern and inter-relation of this object to gain critical information about objects.
	The Figure \vref{intro.fig.w_wo_objects} clearly shows that street objects are strongly organised (top), to the point where removing this organisation (bottom right) negate the purpose of these objects.
	Defining and retrieving relations amongst objects is an old and multidisciplinary problem.
	\citep{Clementini2008} review related references in linguistic, philosophy, psychology, Geographical Information System (GIS), Image processing and qualitative spatial reasoning.
	They propose a common evaluation framework.
	
	More related to the GIS community,~\citep{Steiniger2007} present a coherent typology of spatial relations applied to cartographic generalisation.
	Their typology is general enough to be applied outside of this field.
	In a recent state of the art,~\citep{Touya2014} describe in great details previous works in the GIS field and propose a new taxonomy along with several use cases to illustrate the relations. 
	
	Extracting such relations is a difficult problem, and could be related to Extrinsic/intrinsic symmetries~\citep{Mitra2012}. The real world relations are fuzzy like in the method used by~\citep{Vanegas2013}, where alignment and parallelism spatial relationship in aerial images are defined in a fuzzy way.
	More generally, complex pattern of objects may need a full grammar to be represented 
	( See Section \vref{sota.approaches.procedural_modelling}).

	\subsection{Texture synthesis}
	When the objective is to get a photo-realist 3D model, a possible strategy is to use real or synthetic images and drape them on a geometric street model. 
	This texturing process (or draping) is a major bottleneck for reconstructing a large number of streets.
	Such textures are hard to design, and if using data from sensing, they have to be cleaned.
	\citep{Cornelis2008} uses multi-images to blur the detected vehicles and replace them by detailed 3D model. They also use texture map to efficiently store the road and buildings aspects. 
	A state of the art of texture synthesis and deformation is out of the scope of this article (interested reader could refer to~\citep{Wei2009}).
	To pick a few,~\citep{Cabral2009} uses generic texture deformation by auto-similarity maps to adapt to the geometric deformation.
	Also, although the focus is not the same,~\citep{Ijiri2008} could be used to generate sometime complex pavement pattern of streets ground.

\subsection{Conclusion about street reconstruction}
\label{sota.street.conclusion} 
Street reconstruction is a difficult problem, which is essential for urban reconstruction, but seems to have been much less studied that building reconstruction.
However a large amount of work has been done on road reconstruction, in particular using remote sensing data such as image and aerial lidar. 
However these methods only reconstruct one aspect of a street (geometry, transport related information, street feature, etc.),
and may be of limited precision. 
 
We note that street functions (transportation) and features (objects) are closely interrelated, which indicates the need of a global method taking both into account. \\
It seems that streets are strongly determined by their transportation function.
As such, the role each street plays in the more global street network is a key factor that has to be taken into account when reconstructing this street.
This indicates the necessity to have a multi-scale approach, both at street and street network scale.\\
Street objects have complex organisation (pattern, symmetry), are interrelated, and may also depend on street morphology.
This indicates that a very powerful approach able to model hierarchical patterns is needed. Procedural and/or grammar approach appear to be good candidates for this task.


\section{Street network}
\label{sota.street_network}

In this section we introduce challenges and stakes of street network reconstruction, then propose three classifications of street network reconstruction methods.
The first classification is by the type of road network that is outputted.
The second classification is by the type of input used.
The last classification is by the type of road model used.

Overall, the type of road network output range from simple network to hierarchical network to fully attributed network for traffic simulation.
The input can be from example/template, procedurally (without or with interface), specifying constraints or using GIS data.

Popular procedural methods are L-system, Agent-based simulation, and templates. (\citep{Kelly2006}, page 12).

	\subsection{Introduction to street network reconstruction}
	
	Street network modelling is of particular importance for city modelling.
	A city organisation relies so heavily on street network that it is often the first step of the city modelling process(e.g. CityEngine (ESRI, n.d.)).
	Street are also connected and form a network regulated by traffic laws and many related signs, thus having a specific nature which must be taken into account to enable traffic simulation.
	The street network then becomes a complex graph which exhibit a partial fractal nature~\citep{Frankhauser2008}.
	Reconstructing becomes then much more difficult because the support as well as the connectivity information must be retrieved and coherent.

		\subsubsection{Challenges in street network reconstruction}
		In urban planning designing the properties of the road network is essential for the city growth and for a good interfacing with city surroundings, which makes it a topic of important consequences.
		The financial and environment-related impacts are also enormous (e.g. road network is commonly used to open up neighbourhood, which can significantly increase land price. On the opposite a major urban road can negatively separate a neighbourhood into disconnected pieces, thus weakening the urban fabric).
		The street network is the support of several forms of transport which are entangled. This fact has important repercussions on city reconstruction. A street with major vehicle traffic and bus lanes will be morphologically and functionally very different from a pedestrian street.
		
		Moreover, reconstructing a city without street network would be pointless because the street network is the very object that links every others and assure the connectivity of the urban fabric.
		
		\subsubsection{Why reconstruct street network}
		Reconstructing the street network is essential for numerous applications, being for direct use (navigation), or for indirect information (e.g. gives complementary information about a street that could be used for street morphological properties evaluation for realism or environmental simulation).
		Real world road network maintenance and construction is a massive industry (around 0.5\% of GDP in Europe, according to (\cite{EuropeanUnionRoadFederation2012}, pages 29-30).
		
		At such it is not surprising that major COmputer Assisted Design (CAD) software companies like Bentley\footnote{\url{www.bentley.com/}} and Autodesk\footnote{\url{www.autodesk.com/}} propose advanced products to create/renovate road networks.
		These software features would probably place them at the state of the art, however we could not find enough detailed information to discuss them furthermore.
		
		We note than for procedural city modelling, constructing the street network is often the first task (\citep{Parish2001} and subsequent shape-grammar based city modelling), because street structures the city. 
		
		\subsubsection{Street network and road network}
		Most of the  methods we consider reconstruct road network, and not necessary road network in urban environment, even less street network (that is also reconstruction street features, street objects, etc. ).

		As such, these methods focus on reconstructing a network for vehicle, although streets contains other network, such as pedestrian network.
		Yet pedestrian network can be inferred from road and building (\cite{Ballester2011}), or semi-automatically created with ad-hoc tool (\cite{Yirci2013}) and then updated afterward using GPS trajectories (\cite{Park2015}).
		
		We added methods performing road network reconstruction to this state of the art as they may be applied for streets. 
		
	\subsection{A classification of road network reconstruction methods}
		In this section we propose three classification of street network reconstruction methods: by targeted road network model complexity, by Input type and by road type. 
	
		\subsubsection{Classification by targeted road network model complexity}
		We propose a first ordering of related article by the type of road network they output.
		\paragraph{Flat road network}
		Some methods are suited to design flat road network~\citep{Applegate2011, Galin2010, McCrae2009, Merrell2011}. These roads may adapt to terrain geometry and/or constraint (lake, slope, forbidden area), modelling if necessary bridges, tunnel, over passes, etc. 
		\paragraph{hierarchical road network}
		Yet road network is intrinsically hierarchical (motorway, primary way, etc.), procedural methods are particularly adapted for this.
		For instance, ~\citep{Parish2001, Chen2008, Galin2011, Lipp2011, Yang2013} use multi scale methods, but usually only consider the graphical aspect.
		
		\paragraph{Road network with traffic information}
		Lastly methods can output a complete road network with full navigation attributes for traffic simulation and/or visualisation~\citep{Despine2011, Wilkie2012}.
		These methods are more focused on filtering, correcting errors, constructing a multi-layer data model ( global topological network, lane network for traffic, geometry+texture for visualisation).
		
	\subsubsection{Classification by Input type}
		Methods about street network reconstruction can also be ordered given their data input type.
		
		First some methods directly reconstruct road network using results from sensing, such as aerial Lidar (\cite{Wang2013}), aerial image and radar (\cite{ChuHe2013}), GPS traces (\cite{Ahmed2014,Kuntzsch2015}), or even mobile mapping (\cite{Mueller2011}).
		
		\citep{Merrell2011} is example-based (a given model is analysed, then extrapolated to bigger model).
		A template also plays a role in~\citep{Parish2001} to determine the global pattern of road configuration (e.g. dominant grid-pattern as Manhattan, or dominant radial pattern as Paris).
		Similarly, templates are used for high level road network configuration in~\citep{Yang2013}, but more importantly to design minor roads (indirectly).

		The principal weakness of procedural generation is control (See Section \ref{sota.approaches.procedural_modelling}). 
		Thus many methods try to deal with this by providing interfaces. In~\citep{Applegate2011, McCrae2009} user directly sketches road path in 2D and a 3D clothoid is fitted to the correct elevation and the land is properly dug. In~\citep{Lipp2011} a user directly edits the network graph with advanced operations (copy-past, insertion, rotation, translation) that preserve the graph properties.~\citep{Yang2013} propose some control via constraint layers (e.g. a lake surface, or a given type of organisation for an area).
		
		Similarly, many methods use constraints as input. Typically mechanisms permit to define area where road network is constrained, for instance in parks and/or river.~\citep{Chen2008, Parish2001, Lipp2011}.
		
		It is different for~\citep{Galin2010, Galin2011} where the constraints system is at the hearth of the method.
		In these articles, custom cost functions, special constrains (park, highway without intersection) and a specialised solving system allow the system to generate an optimal path for the road taking into account the geometry and the nature of the terrain (constructing bridges or tunnels along the way).
		
		Input data can be even more abstract as in~\citep{Despine2011, Wilkie2012}, where they use polylines with attributes.
		The challenges are then as much to filter and correct input as to use methods to generate a complete road network data suitable for traffic simulation.
		
		\subsubsection{Classification by geometry representation of roads}
		We can also classify the methods for street network reconstruction by the way they model the road surface. 
		
		The clothoid is a popular way to model road. This is due to the fact that clothoids are mathematical curves along which curvature varies linearly, thus conducting to a pleasant acceleration while driving. Dut to this property, clothoids have been used by civil engineer for a long time to construct actual roads.
		Clothoid can be extended to piece-wise clothoid or super-clothoid~\citep{Bertails-Descoubes2012}. 
		In urban environment, acceleration constraints are often less important that historical heritage or global city layout, thus the model iseems to be less used.

		Another popular parametric model is based on arcs (circular arc:~\citep{Wilkie2012}, parabolic arcs:~\citep{Despine2011}, or just polylines:~\citep{Parish2001}).
		See (\cite{Wilkie2012}, pages 2-3) for more geometric primitives for road modelling.

	\subsubsection{Other methods that could be applied to road network reconstruction} 
	Some methods are adapted to model 3D network like a road network but where tested on other fields. 

	For instance~\citep{Merrell2011} is a very general procedural modelling method that analyse an input shape (geometric constraint) in order to create a new bigger procedural model, respecting some user defined constraints.
	In a different direction,~\citep{Krecklau2011} propose a custom grammar adapted to interconnected structures.
	By defining potential attachment points, and geometrical queries able to find potential connections, there grammar allows to model different kind of interconnections.
	This may be naturally extended to road network modelling, taking advantage of the connectivity that defines a road network.

	\subsection{Conclusion}
	\label{sota.street_network.conclusion}
	The street network is essential for urban reconstruction, as it defines many aspects of the city, and is paramount in the way streets are used. 
	Even more important, the street network is a structuring element for a city, similarly to how the street axis is structuring for street.
	This indicates that an urban model could be based upon the street network.
	
	Most methods focus on road network, few consider urban environment, and no method reconstruct a real street network, including pedestrian network, and vehicle network. 
	In the same spirit, not all methods produce a hierarchical network, even fewer with geometry and traffic information.
	
	The difficulty seems to be coming from the fact that a streets network is a graph embedded in 3D, which makes it much more abstract than the sensing information, hence the complexity.
	In particular, the intersections, bridges, tunnel, fly-over are supplementary difficulties. 
	
	Because o  this complexity, many methods have to rely on user inputs.
	This indicates that having an interactive editing capabilities of the street network is important and necessary.
	
	Some aspects of street network are impossible to determine without street features.
	For instance the number of lanes has to be inferred from markings, the connectivity of the network from traffic lights and traffic signs, etc.
	This seems to indicate that a street network reconstruction has to be done at two scales: at the network scale and at the street scale.


\section{Urban vegetation reconstruction}
	\label{sota.vegetation}
	In this section, we introduce why reconstructing urban vegetation is an important part of urban modelling, and which challenges it creates.
	We then discuss vegetation reconstruction and the various strategies and scale at which it can be done, then we propose three classifications of vegetation reconstruction methods.
	
\subsection{Introduction} 
	\subsubsection{Why reconstruct vegetation in urban area }
		\paragraph{Vegetation plays an important role for city}
		The vegetation has been primordial for Mankind for a long time.
		Forests occupy a large part of land surface (30\% in France).
		It is then not a surprise that the vegetation is very common and plays a very important role in cities.
		
		The vegetation in cities has a significant influence on noise propagation, air quality and temperature, water cycle, and also has a significant impact on human social behaviour.
		Each of these aspects covers a vital part of urban planning, be it for comfort (temperature, air quality, human behaviour), or for technical management advantages (water cycle, noise, wind).
		
		\paragraph{Vegetation play an important part for city modelling}
		In a pure 3D reconstruction, the vegetation is important for realism and because it is geometrically so different from its surrounding (a sparse organic spherical form, as opposed to the locally planar and compact rectangular form of buildings or streets).
		As such, methods devised for buildings reconstruction are usually sub-optimal at best for tree modelling.
		Because trees are large and recognisable, they alter much the perception of a street. 
	
		\paragraph{Vegetation is very present in city}
		As a numerical example, about 5\% of Paris surface is dedicated to parks, that is not taking into account the two small forest that are officially within Paris (bois de Vincenne and Bois de Boulogne).
		The number of trees in streets is above 250 000 in Paris.
		This means that in average there are trees every few dozen meters in Paris streets.
		
		\paragraph{Vegetation reconstruction is useful for other methods}
		Even when reconstruction of vegetation is not explicitly wanted, it can be of great help to have a vegetation model (possibly implicit) for reconstructing other objects occluded by vegetation.
		Another important use of vegetation is for landmark maps.
		In such context the large visual space a tree occupies is precious because it is easily recognised.( See~\citep{Soheilian2013} for a state of the art  of landmark based localisation, and ~\citep{Brenner2010} for a localisation using exclusively trees).
	
	\subsubsection{Challenges in urban vegetation reconstruction} 
			
		Reconstructing the vegetation in an urban environment is challenging for several reasons, some due to the nature of the vegetation (multiscale, ecosystem), some more technical (sensing data precision and completeness, scaling).
		
		\paragraph{Vegetation is a multi-scale complex ecosystem}
		Vegetation is often a whole ecosystem, with several species living together.
		Like many living organisms, plants exhibit a fascinating multi-scale nature with fractal-like properties. 
		
		Therefore one must define up to which scale the reconstruction process should stop.
		
		To the best of our knowledge the current state of the art for trees is at the branch scale, with reconstructed trees having a similar leaf organisation as the model~\citep{Pirk2012}, but not an exact leaf to leaf reconstruction.
		However a recent work on small plants suggests a future move toward the leaf scale~\citep{Li2013}.
		Concerning the vegetation reconstruction, most of the works reconstruct the vegetation in the form of a distribution of species.
		
		\paragraph{The large number and scaling challenge}
		The vegetation uses large amount of city surface, and in streets each tree may occupy a large volume.
		Moreover, the scaling problem is evident when considering that each tree may have hundred of branches, and there are hundred of thousands of trees.
		
		\paragraph{Modelling trees at city scale} 
		At the city scale, a specific modelling strategy may be needed for trees,
		as any use of the tree models must introduce a reasonable hardware requirement.
		To this end, a solution is to have several models for the same tree with different level of details.
		
		Trees have a fractal nature, which can be leveraged to allow the efficient modelling of large areas with many trees.
		A less detailed model can be rendered when the tree is far from the viewer, while the more precise model is showed when the tree is close.
		\citep{Livny2011} produce different levels of details for every model of trees. 
		Similarly, the popular XFrog\footnote{\url{http://xfrog.com}} can also be used to produce levels of details. 
		When the trees are regrouped, one could also rely on tailored methods to allow realistic and fast visualisation (e.g.~\citep{Bruneton2012}).
		
		\paragraph{Vegetation is hard to sense}
	
		Another point is that trees are by nature occluding elements from an aerial point of view. 
		This stems from the tendency of the trees to capture sun light coming from above, hence they limiting the picturing.
		For this reason, removing the trees for correct façade reconstruction is a very classical problem in terrestrial laser and image processing. 
		
		Tree reconstruction is also challenging because the sensors (image, Lidar) give information about surfaces, which is fine for a building, but may fail to pass the tree crown to get the branching structure (full wave or hyperspectral Lidar somehow mitigates this).

\subsection{Vegetation reconstruction} 
	A global state of the art on vegetation modelling and reconstruction is out of scope of this work; therefore we will only give an overview of vegetation modelling and focus on its use in urban context.
	We also included some method for tree modelling, as these could potentially be used for tree reconstruction using an Inverse Procedural Modelling paradigm.
	
	In this section, we briefly introduce the strategies for vegetation reconstrution, then consider the different scales to which the vegetation can be reconstructed.
	We then propose three classifications of methods for vegetation reconstruction.
	
	\subsubsection{Strategies for vegetation reconstruction}
		
		\paragraph{Focussed on trees}
		Vegetation reconstruction usually focuses on tree reconstruction, even though some methods output an ecosystem type rather than a tree species (\cite{Gong2002}).  
		Orthogonally new Lidar technologies allow accessing more tree properties.
		For instance~\citep{Hakala2012,Wallace2012} recover tree properties and canopy properties using multi spectrum Lidar technology.
		
		\paragraph{A more model oriented reconstruction}
		The strategies for vegetation reconstruction are slightly different from typical strategies for man-made objects reconstruction (including buildings and façades).
		This is due to the fact that tree species evolve slowly and have been known for centuries, along with key properties of each species.
		Moreover, urban tree species are much more limited (order of magnitude : 100) than potential street furniture types for instance (order of magnitude : 10000).
		
		For these reasons, and because of the multi-scale problem, top down approaches (model oriented) seems to be much more popular than bottom up approach (data oriented).
		That is, most model have strong hypothesis and model which are fitted to sensing, rather than directly using sensing to reconstruct trees from scratch. 
		
		This is quite different from building reconstruction, where building styles can be mixed, and each building does not necessarily fully enforces a style.
	 	
		\paragraph{Tree reconstruction or tree growing}
		There are two main approaches to reconstruct the vegetation in cities: an analytical approach, where we try to retrieve direct morphological information about the tree to reconstruct it as is, and a more synthetic approach where we try to retrieve general information about the tree (species, height, crown seize), then synthesize it using growth model and known parameters of the species.
		
	\subsubsection{Choosing a scale for reconstruction}
		Vegetation is multiscale, therefore, before reconstructing, the targeted level of detail has to be chosen.
		\paragraph{Forest}
		Forest management is a century old tradition. 
		Therefore forest models have been developed, such as group of trees species repartition, possibly with their age, height, crown size, etc.
		These are used for forest exploitation, land planning and so. 
		Such models are commonly obtained by field surveys, along with information obtained from remote sensing technologies (aerial images, Lidar)~\citep{Gong2002}.
		For example~\citep{Watt2013} use full-wave ground Lidar to estimate two exploitation-related characteristics of a patch of forest.
		\paragraph{Patch of trees}
		It is also common to model homogeneous patches of ecosystem, with a larger scope than tree alone, sometimes involving plants modelling. This allows height/species/spatial statistical distribution analysis.
		\paragraph{Individual tree}
		Tree models have been actively researched, including tree growth characteristics and species specificities. Procedural modelling methods are especially popular.
		
		\paragraph{Individual plant} 
		Plant modelling is also an age old tradition \citep{VanGogh1888}, with many applications in design and entertainment. 
		More recently plant reconstruction has also been tackled (\cite{Li2013}).

\subsection{Classifications of urban vegetation reconstruction methods}
		We propose three classification of methods related to vegetation reconstruction.
		
		\subsubsection{Classification by input data type}
		We classify the vegetation reconstruction method based on the input they use, from dedicated Lidar to more generic remote sensing, to interactive feedback (human interaction).
		
		Input data for tree modeling can be point clouds from Lidar tailored acquisition~\citep{Preuksakarn2010,Livny2011} or general acquisition~\citep{Livny2010}, as well as point clouds from dense matching~\citep{Li2013}.
		Some methods also use aerial images~\citep{Iovan2013}, or semantic maps~\citep{Benes2011}.
		Some methods are based on constraints on the tree growth~\citep{Pirk2012,Runions2007, Talton2011}.
		Lastly, many methods rely on user feedback but may be automated by using remote sensing data inputs~\citep{Krecklau2012, Krecklau2010, Lintermann1999}.
 
		\subsubsection{Classification by modelling method}
		We propose another classification of tree reconstruction following the modelling method they use,
		from procedural methods to L system to generic grammars.
		
		Individual tree modeling is a mature research interest. It has been historically focused on procedural methods. Mature interactive commercial solutions such as XFrog~\citep{Lintermann1999} exist and are widespread.
		In most cases the trees are modelled procedurally, possibly using parametrised shapes like generalized cylinders~\citep{Bloomenthal1985,Li2013,Pirk2012, Preuksakarn2010,Xfrog2014}.  
		
		Explicit grammar systems are also popular, in particular the L-System grammar~\citep{Deussen1998}. More general grammars have been extended to produce trees along with more rectangular objects ~\citep{Krecklau2012, Krecklau2010}.
		See Section \ref{sota.approaches.procedural_modelling} for more details about procedural modelling.
	
		\subsubsection{Classification by Reconstruction strategy}
		The last classification of the reconstruction methods we propose is by reconstruction strategy,
		from direct from data, to analyse-synthesis to inverse procedural modelling to whole urban ecosystem design.
		
		Reconstruction strategies can be straightforward~\citep{Livny2010, Preuksakarn2010} from direct remote sensing data. It requires however high quality data and has not been experimented on city scale.
		However the reconstructed trees can have similar look and properties as the real one up to the level of group of leafs~\citep{Livny2011}, or even the leaf level~\citep{Li2013}.
		
		Other methods focus on an analysis-synthesis approach. 
		The goal is to retrieve a number of properties of the tree (species, height), along with constraints introduced by its surrounding, then use a realistic growth method to obtain a tree model hopefully close to the real tree.~\citep{Runions2007} use a space constraint approach to model the competition for space, while~\citep{Talton2011} constrain the tree leaf coverage by a bitmask, and~\citep{Pirk2012} add solid object constraints as well as shadow influence.~\citep{Iovan2013} use images to detect and classify urban trees, then use the extracted parameters as well as space constraints to grow plausible urban trees.
		
		~\citep{Benes2011} are even more generic and introduce man-related constraints on a city area: in some part of the city vegetation growth is strictly controlled (trees species and spatial repartition), in other the control is less strict. Trees are also spreading over time. The system is then evolved over a period of time to generate 3D space + time tree repartition and visualisation.

\subsection{Conclusion for urban vegetation reconstruction}
\label{sota.vegetation.conclusion}
Vegetation is important for city modelling, both by its sheer presence, the roles it plays (temperature, pollution, noise, water, human perception, etc.), and its interest for urban modelling (street morphology, occlusion, landmark for registration).\\ 
Yet, the vegetation is hard to reconstruct (complexity, multi-scale, volume), and most methods focus on trees.
 
Because vegetation exhibit a regular and hierarchical nature, procedural modelling methods seem to be very indicated.

We note that the vegetation strongly depends on other urban features. 
Plant species will be influenced by the typology of area (residential, industrial, etc.), plant growing will be influenced by buildings, and realistic trees will most likely be pruned, therefore being influenced by road surface, and some road feature (road surface, traffic light, traffic sign).


\section{Urban features}
	\label{sota.urban_feature}
	\subsection{introduction to urban feature reconstruction}
	We consider only man-made urban feature reconstruction (See Section \vref{sota.vegetation} for vegetation).
	We found few methods dedicated to urban feature reconstruction (street furniture, markings, etc.).
	Therefore we also integrate generic methods for man-made object reconstruction in this state of the art. We consider that these methods could also be applied on street objects.
	
	Urban features comprises urban furniture (e.g. barrier), markings (e.g. lane separator markings), but also features of the street such as local height of sidewalk limit, etc.
	 
	\subsubsection{Importance of street features for city}
	A city contains large amounts of street features, such a street furnitures, markings, etc. 
	These are important by their number (over 1 million in Paris), by their diversity (over 13000 references on a site like \citep{ArchiExpo2014}), and above all by the functions they fulfil (information, security, decoration, etc.).
	Street furnitures are seldom randomly placed and chosen, but instead are essential tools for the complex social interactions that a city host.
	Figure \vref{intro.fig.w_wo_objects} shows well how position and relations are important for urban features.
	
	Virtually any human behaviour in a city relies on street objects, essentially because street objects regulate transport (information, rules, isolation, whatever the modality) and play a role into managing the city (waste, water collection). 
	
	\subsubsection{Importance of urban feature modelling}
	Modelling urban feature is then essential for traffic simulation, and also for realism (some piece of street furniture have achieved a landmark status, like Curitiba bus stations\footnote{\url{https://en.wikipedia.org/wiki/Rede_Integrada_de_Transporte}} in Brazil).
	Street furnitures can also be extracted to form a landmark map, thus assisting in the georeferencing of a vehicle or user with basic sensors (\citep{Hofmann2009}).
	Street feature also strongly separate urban space (between sidewalk and roadway for instance).

	\subsubsection{Challenges for urban feature reconstruction} 
	Reconstructing urban features is difficult because of their relatively small size, essentially disabling any air sensing, and making it difficult to have precise and complete data (e.g., only a part of a parking meter would be on a street view or on a Lidar acquisition). 
	The geometrical complexity may be high or deceptively simple (e.g. traffic signs are almost pure 2D). The material used can also complicate data sensing (glass, shiny metal, reflective paint).
	However such man-made objects typically expose strong regularities, symmetries, as well as a dominant plan-based structure which can be used by methods to improve reconstruction.
	
	\subsubsection{Reconstructing urban features}
	As always in a reconstruction problem, we have to define up to which scale the objects are to be reconstructed.
	For instance when reconstructing a street bench, shall we simply reconstruct the bench type and orientation, or shall we reconstruct it as several plans with texture, or shall we reconstruct each plank composing it, or shall we even reconstruct how the plank were bolted together, etc. 
	
	It seems that this level of reconstruction is dictated by the quantity and precision of input data, as well as how much the method is model driven.
	This problem is especially pregnant in streets, were the most precise data (order of magnitude of 0.01 \metre) are limitating, as well as the large occlusions.
	
	Of course this level of reconstruction also depends on the intended applications, a proper generalisation is often necessary for performance reasons (trying to render the nails in the hundred of thousands of Paris street furnitures would most likely fail and be useless).


	\subsubsection{Input types}
	
		Traditionally street feature reconstruction methods use street lidar and images \citep{Golovinskiy2009,Soheilian2013}.
		In the more general object reconstruction field, other methods use noisy point cloud from images or color and depths devices (RGBD camera, like the Kinect) (\citep{Stuckler2012}). 
		Even farther, some methods directly use 3D models (\citep{Shapira2009}) to analyse structure and match it against a database. 
		Some methods inputs are even more abstract, like a set of relations among objects (\citep{Yeh2012}), or interactive user inputs (\citep{Gal2009}).
				
	\subsubsection{Hypothesis on street features} 
		Street objects will most likely be severely occluded during sensing. 
		Therefore, making hypothesis is necessary.
		For many methods the hypothesis are to exploit regularity of man made object by using combination of simple geometric primitives (plane, sphere, cone, cylinder ...) with strong common properties (e.g planes will tend to be parallel or orthogonal, axes of primitives will tend to be collinear ), and symmetries.

		\citep{Lau2011,Umetani2012} add another level of constraint by stating that the object can be fabricated (e.g joins between parts must have adequate resistance and the global object must be stable). 
		On another level \citep{Yeh2012,Grzesiak-Kopec2013} use relationships between objects to define constraints that the reconstructed objects must satisfy.

	\subsubsection{Strategies for urban feature reconstruction} 

		Because precise street feature reconstruction is quite new and connects to many research communities, we include methods with very different inputs which could be used for street feature reconstruction, even if not explicitly stated by the corresponding articles. 
		
		Some methods reconstruct directly street features (low level reconstruction), but the sensing data is sparse and often of relative low quality considering the scale of the considered objects.
		As it is often the case a way to simplify a problem too wide is to add constraints and knowledge about it. 
		Some of the approach therefore add strong hypothesis about the object to reconstruct (Section \ref{sota.urban_feature.low_level}). 
		
		Because reconstructing directly street feature may not be feasible, some approaches turn to classical segmentation/classification methods (Section \ref{sota.urban_feature.object}).
		
		This allows to decompose the reconstruction problem: First find which street object is where, possibly determining some of its properties, such as its orientation. 
		Second, find or generate a similar 3D model and populate the reconstructed street with it.

		However, finding the exact corresponding model from incomplete data for a street feature may be challenging (see introduction of this section). 
		Therefore other methods are based on object structure analysis, decomposing it into parts. The reconstruction is then facilitated by the possibility to switch parts of the object as well as complete missing parts by a similar one (Section \ref{sota.urban_feature.structure_analysis}). 
		
		Another more radical approach, which we could call extreme classification, relies on an extensive catalogue of objects. The reconstruction process amount then to find the model in the catalogue that is the closest to the sensed object, then use the catalogue model as the reconstruction.
		
	\subsection{State of the art}
			
		\subsubsection{Low level reconstruction}
		 \label{sota.urban_feature.low_level}
			
			\paragraph{Intro}
				There is a great body of literature about generic surface reconstruction, bet it flat or curved. A naive approach could be to use these methods to directly reconstruct the street objects.
				However due to the massive amount of occlusion (a street feature is commonly occluded halfway), strong hypothesis about the object nature are necessary.
				Also, these methods do not provide semantic information about the reconstructed object (e.g a reconstructed poll wont be identified as a poll but as a cylinder).
			
			\paragraph{Direct surface reconstruction}	
				
				\citep{Bessmeltsev2012} propose a method to directly generate surfaces from 3D lines as input. The extreme data sparsity is similar to what may be available in street feature reconstruction. The authors interestingly make an hypothesis about what type of surface could be expected from a man-designed object.
				
				Using a noisy point cloud \citep{Guillemot2012} make hypothesis on repetitions in the data to reconstruct a better surface. Their method defines local patches as small set of points. When reconstructing the surface of a patch they use the local information as well as informations of similar looking patch elsewhere in the point cloud.

			\paragraph{Simple geometric primitives}
				With dense noisy point clouds of man made objects,
				\citep{Li2011} assume that an object consists of regular geometric primitives globally aligned.
				So, they iteratively detect the primitives with the associated points that support it.
				Then they extract and enforce global relations among these primitives and remove the associated points from point cloud, before iterating on the reduced point cloud. 
				(\citep{Labatut2009,Lafarge2013}) propose other primitive-based approaches applied to buildings which may be transposed to street features reconstruction. The goal of the two works is to extract a mixture of geometric primitives and free-form mesh from noisy stereo-based point clouds.
				One relies on a binary space partition tree and a RANSAC detection method while the other uses a sophisticated energy-based Jump-Diffusion process.
			
			\paragraph{Shape grammars}
			
				The shape grammars like the one defined by \citep{Krecklau2011} generalise the simple geometric primitives.
				They are by construction well adapted to represent man-made objects (and even vegetation (Section\ref{sota.vegetation})
				Such grammars have a great generative power, but one has to resolve an inverse problem to use them for reconstruction. 
				
				This problem is solved via \myemph{the Inverse procedural Modeling methodology} (See Section \ref{sota.approaches.inverse_procedural_modelling}).

		\subsubsection{Object reconstruction}	
		\label{sota.urban_feature.object} 
			\paragraph{Introduction}
				Given the occlusion in data, it may not be possible or satisfactory to reconstruct objects directly. Therefore many methods chose a two steps approach, where the first step detects and classifies objects in the input data. 
				The second step can then be adapted to each object type. For each object type the options are either to reconstruct it directly using tailored methods or to populate the street with a model of this object.
				
				Compared to low level reconstruction Section \ref{sota.urban_feature.low_level}
				, these methods can be fitted to each objects, and the inserted models are cleaner than model reconstructed from scratch.
				A complete example of this workflow is given by \citep{Cornelis2008}. They use video streams from a street vehicle to reconstruct a 3D map of a city. Along the way they detect cars on the side of the road (3D bounding boxes).
				Ultimately, they insert into the 3D city model clean 3D car models in these bounding boxes.
				This greatly improves accuracy of reconstruction and realism of city model.

				Classification is a transverse problem in many computer science fields. Street objects classification must be adapted to challenging input data (scale, occlusion, sparse data). Also, as stated in the introduction of this section, the number and types of street features is important.
				This proves to be a major obstacle for machine learning methods which rely on training data.
				In these training data some uncommon objects may be statistically overwhelmed by more common (and similar) objects (see \citep{Golovinskiy2009}).
				
				Another set of difficulties is added by the second step, which imposes not only to classify objects, but also to measure parameters to correctly insert models (orientation, state, potentially more parameters for parametrised objects).
				
				We order the related methods by the detection / segmentation / classification task, the feature extraction task and the matching task. Such order is only practical because many methods mix these categories.
				In classification literature the word feature is often used instead of descriptor. We choose here to use the word descriptor to not introduce confusion with the topic (Urban/street feature reconstruction).
				
				
			\paragraph{Detection, segmentation, classification}  
		 
				In an influential article, \citep{Golovinskiy2009} use street Lidar input to demonstrate the full localisation/segmentation/classification pipeline. They test multiple classifiers methods and descriptors, and perform an experiment on large scale real world data. Their method detects around twenty different object types.

				\citep{Shao2012} also illustrate a full pipeline but not in a street object context. They use interactively segmented colors and depth images (RGBD). The extracted objects are then matched against a database of 3D models. These models are inserted using an optimisation process to determinate their size and position.
			 
				
				In order to tackle the scale problem, \citep{Yu2011a} propose a segmentation of a massive city point cloud into ground and façades, and potential objects.
				The work of \citep{Lippow2008} adapt to the many type of objects to detect (in the computer vision field).
				Their method learns an AND/OR probabilistic tree for a category of object in annotated images. Such trees are then used for detection, not of one object, but of the category of this object.

				
				Some usages do not necessitate accurate object reconstruction. For instance \citep{Hofmann2009,Soheilian2013} detect poles (respectively streets signs and markings) based on simple geometric model in order to create a landmark map which can then be used to cheaply localise other data. 
				
				\cite{Timofte2011} focus on manhole detetection and reconstruction using a mix of 2D and 3D methods for image processing.
				
				Similarly, building in real time such localisation map with 3D semantic voxels \citep{Stuckler2012} significantly improves the registration of their colors + depth images data (RGBD). These voxel maps may also be used for more abstract task like human-robot communication.

			\paragraph{Descriptor extraction}
				 
				The task of classification is often very sensitive to the choice of descriptors of an object. A good descriptor should reduce the amount of data necessary to describe the objects, but not reduce the information much.
				Furthermore, the descriptors must be chosen to be differentiating between object types. A good choice of descriptors increases recognition rate and reduces errors.
				 
				We refer to the appropriate articles for the classical descriptors used by \citep{Cornelis2008,Stuckler2012,Shao2012,Soheilian2013} (Implicit Shape Model, simple local RGBD  descriptors, many descriptors selected through Random Forest, image and contour-based).

				Concerning the shape-matching methods, the choice of descriptor is of the essence. The method performance, speed, scaling and accuracy strongly rely on it.				
				\citep{Papadakis2007} use descriptors based on spherical projection, \citep{Shao2011} use depth feature as well as geometric primitives, \citep{Eitz2012} use adapted Gabor filters.

				For noisy point cloud data, The work of \citep{Kalogerakis2009} who extract lines of curvature may also be used as a descriptor for street feature. According to the authors, this curvature-based descriptor is specific to man made objects.

				\citep{Golovinskiy2009} outline that contextual (i.e. relational) descriptors are of great use for object classification. 
				
				In that way, \citep{Vanegas2013} propose a fuzzy relational descriptor that may be adapted to noisy and incomplete data. Using aerial images, the proposed method extracts fuzzy spatial relations between objects like alignment and parallelism.
				 
				A very complete generalisation of these kinds of relationships is given in \citep{Mitra2012}. This state of the art provides numerous useful reflections about the presence of total or partial symmetry in man made objects.
				For example, \citep{Xu2012} propose a method to compute partial symmetries at multiple scales. 
				Such relations could be used as high level descriptors.

			\paragraph{Model matching}
				 
				To the best of our knowledge no matching system against a 3D model database has yet been applied to street feature reconstruction.
				However such systems have been developed in the field of model matching. These methods may be transposed to the field of street feature reconstruction, as demonstrated by \citep{Shao2012} for indoor objects.
				In fact, most of the presented shape matching methods use 2D sketch produced by a user.
				Nevertheless such an input could be conceptually replaced by the sensing data of street feature.
				 
				The pipeline of \citep{Papadakis2007,Shao2011,Eitz2012,Shao2012} is similar and can be decomposed into an off-line data base creation step, and an on-line query step.
				First the methods extract descriptors for thousand of 3D objects and constitute a database associating object model with their descriptors. 
				During the on-line step, an user input of a 2D drawing is analysed, the same descriptors are computed and the methods search the 3D models in the database that have the closest descriptors to the user input. The result is a list of matching shapes from database, with a matching score. 
				
				Howsoever these methods differ by the choice of descriptors, the validation (\citep{Eitz2012} analyse the best way to perform dimension reduction (i.e. translating optimally a 3D model into 2D views)), and the reconstruction step (only performed by \citep{Shao2012})).
				
				\citep{Jain2012} also perform shape matching, but in a fundamentally different way. The goal of the author is to automatically transfer materials (i.e. texture, colours and lightning) to a 3D model by matching its different parts with a 3D model database.
				The authors also follow the two steps that are the constitution of a database of 3D models, and then a query step.
				The originality is that the database is a graph of parts of models that is automatically computed based on similarities of parts (spatial, geometrical, material-wise). 
				Querying the database then amounts to compute the graph for the queried 3D model, then add this graph to the database graph and use a loopy belief propagation algorithm to perform inference.
				
				Interestingly such method introduces the use of structural information about objects. This information is pivotal to estimate the material of each parts.

		\subsubsection{Object structure analysis}
		\label{sota.urban_feature.structure_analysis}
			Man made objects are constituted of parts having (potentially hierarchical) relations (symmetry, fixed angles, etc). This relations describe the object structure.
			
			\paragraph{Intro}
				Object structure analysis may be of great help in street feature reconstruction, and this at two scales. At the part scale (decomposing an object into structured parts, e.g. a street light may be a cylinder (pole) and a sphere (light bubble)), and at the multi object scale (decomposing multi objects into structured objects e.g. a dashed marking line may be described as a repetition of aligned small pieces of plain lines.)

				Such structure analysis may be useful at the object scale, because analysing the redundancy, structure and organisation of an object allows to extract higher level information about it.
				It can then be used to compensate noisy or incomplete data (\citep{Shen2012} do this in a reverse way)(e.g sensing only the front part of a pole may be sufficient if we have the information that poles follow a rotational symmetry).
				
				Moreover, a strong structural information and presence of symmetries \citep{Mitra2012} is typical of man-made objects and may be used as descriptors for classification/matching~\citep{Shapira2009}. 
				Alternatively, such regularities allow for compression and Levels Of Detail (\citep{Jang2006}).
				This also gives an information orthogonal to pure geometric comparison: it allows to measure how similar the structure of two objects is, rather than their geometry. For instance, a motor bike and a bycicle are structurally similar, but may have very different geometries. 
				
				Secondly, some methods that leverage structure of object may be generalised at the multi-object scale, i.e. finding and using the structural relations between objects, that are known to structure the layout of objects in a street ( See Section \ref{sota.street}).

			\paragraph{How to detect symmetries}
				
				Analysing the structure of a 3D object is complex because it involves unsupervised segmentation as well as a relation extraction between parts.
				
				Among the relations used in the methods (generalised), similarities are popular.
				 
				A typical approximate symmetry pipeline is given in  \citep{Mitra2006}, where the input is a 3D model (which could also be a dense 3D pointcloud).
				In a first step they get random sample points from the surface, and perform pairwise symmetry parameters estimation by taking into account a patch around the points. \\
				Then, in the space of the found pairwise-transformation, a clustering is performed to extract dominant transformations.
				The supports for this transformation are then computed by region growing from the sampled points. 
				
				\citep{Xu2012} improve this process by adopting a multi-scale classification.
					
				The authors of \citep{Li2011} choose another direction and perform the equivalent of relation clustering with a custom graph simplifying algorithm.
		 	
				Whereas partial symmetries are covered in \citep{Mitra2012} as a generalised case of symmetries,  \citep{Vanegas2013} incorporate them in the fuzzy logic theory.
				Exploiting ad-hoc fuzzy operators, they propose a way to compute fuzzy parallelism, fuzzy alignment, etc.	
				 
				The work of \citep{Cullen2011} generalise more the symmetry concept by constructing a tree of symmetry compositions representing a pattern. This method is  very close to a procedural expression.
				After having computed such trees for two patterns, they can be merged to create a hybrid pattern that mixes the two input patterns.
		 
			 
			 	Other methods use touching relation to extract structure.
			 	
		 		\citep{Shapira2009} use a custom descriptor based on local diameter of the object. They use it to iteratively fit Gaussian mixtures in order to find parts, then build a graph representing the relations between parts. They can then perform parts matching taking into account the context of the parts to match.

		 		\citep{Jain2012} extract structure by contact and symmetry analysis, and use it for matching or for generation of hybrid models by genetic evolution.

		 		\citep{Lau2011} retrieve an even more complex structure as they perform inverse procedural modelling (See Section \ref{sota.approaches.inverse_procedural_modelling}). 
		 		They analyse contacts between parts of an input 3D model, parsing it into a graph of connections. Then they use a custom grammar to express this graph by inverse procedural modelling. Using the grammar with the extracted rules and parameters generate a fabricatable 3D model.

		 		For completeness sake we mention that some methods consider the decomposition of object into parts as a preprocess step that has already been performed (\citep{Chaudhuri2011,Xu2012,Shen2012}).

			\paragraph{Using the structure} 
			 
								
				At the object scale having such a structural description of objects allow \citep{Shen2012} to match parts of 3D model on noisy and sparse RGBD point cloud. 
				The authors of \citep{Chaudhuri2011} tackle another problem by suggesting parts when building a 3D model from scratch. Yet their method may be used to complete occlusions resulting from street feature sensing.

				Expressing the object structure is not necessary if the goal is to respect symmetry relations between parts of 3D models when editing (structure preserving editing).
				For instance \citep{Gal2009,Bokeloh2011,Bokeloh2012} analyse a 3D model to detect symmetries (respectively more general patterns), which produce a set of constraints that are linearised, allowing to edit the shape interactively while computing a solution respecting the constraints by propagating the changes and minimising an energy locally (respectively minimising an energy).

				
				At the multiple objects scale, \citep{Krecklau2011} propose an extension to their grammar that add the possibility to model interconnected structures, which are common in street.
				However using such grammar would require to extract relations and patterns amongst street objects.
				In the same field \citep{Grzesiak-Kopec2013} adapt a shape grammar to resolve a layout problem.

				Other methods uses these high level data that model the relationship between objects. Still, in all the article we present these relational data are user input and are not extracted (with the exception of \citep{Fisher2012}).
				 
				Putting in leverage these relations allows to use powerful optimisation methods to generate a good placement for furniture in a room in (\citep{Yu2011}).
				One limitation is that the number of objects is fixed.  
				 	
				\citep{Yeh2012} remove this restriction by proposing a similar method that uses another advanced optimisation framework to find conjointly the number of objects, as well as their position and orientation. 
						
				Those two methods could be used in street object reconstruction by resolving an inverse problem : given noisy observations and relations, find an optimal objects reconstruction.
				
				Interestingly, \citep{Fisher2012} directly extract relationships between objects from a clean 3D indoor scene using Bayesian networks and Gaussian mixtures. In a further step they generate a new scene with objects matched from database satisfying the extracted relationships.
				
				The relationship between street features is discussed in detail in the section 
				\ref{sota.street}.

\subsection{Conclusion}
	\label{sota.urban_feature.conclusion}
	Urban features are important (number, role).
	Urban features are strongly dependent on context (a same white marking could have totally different meanings if it were on the road surface or on the sidewalk).
	Reconstruction is difficult because data is sparse, yet because the objects are man-made, many hypothesis can be made.
	When this is not sufficient, user interaction is necessary.
	Many reconstruction strategies are possible, from direct reconstruction, to model oriented reconstruction, to procedural modelling and grammar, to use of catalogues of objects.

\section{Conclusion}
\label{sota.conclusion} 
In this chapter we tried to consider all aspects of urban modelling/reconstruction (street, street network, vegetation, urban feature).
Each one of this aspect has a dedicated conclusion ( 
Sec. \vref{sota.street.conclusion},Sec. \vref{sota.street_network.conclusion},Sec. \vref{sota.vegetation.conclusion}, Sec. \vref{sota.urban_feature.conclusion} )

There are common elements for all these aspects of urban reconstruction. 
The first element is that each aspect is important for urban reconstruction. 
We can not simply reconstruct buildings to reconstruct a city, other aspects also have to be reconstructed.

The second element is that reconstruction is difficult for each aspect, the challenges come from the complex nature of urban environnement and from the limitations of available data.

The third element is that all the aspects of urban reconstruction seem to be linked.
Street network reconstruction require information about urban feature, which are influenced by street morphology, which influence urban vegetation.

The last element is that many strategies are available to reconstruct each aspect, from direct reconstruction to procedural modelling.
(Inverse) Procedural modelling seems to have potential to reconstruct all the aspects.

\section{Acknowledgment}
This article is an extract of \cite{Cura2016thesis} (chap. 1). We thank Prof.Peter Van Oosterom and Prof.Christian Heipke for their extensive review.


	\section{Bibliography}  
	\bibliography{./all_bibli} 

%
%
%
%

\end{document}